\documentclass[12pt]{article}
\pdfoutput=1
\usepackage{makeidx}
\makeindex
\usepackage[a4paper]{geometry}
\usepackage{jheppub,amsmath,amssymb,amsfonts,amsxtra,mathrsfs,graphics,graphicx,amsthm,epsfig,ytableau,bm,longtable,float,color,tikz,mathtools,xfrac,footnote,rotating,lscape}
\restylefloat{table}
\usepackage{dsfont}
\usepackage{multicol}
\usepackage{lipsum}
\usepackage{textgreek}
\usetikzlibrary{decorations.pathmorphing}
\usetikzlibrary{decorations.markings}
\usetikzlibrary{quotes,arrows.meta}
\usetikzlibrary{arrows, decorations.markings, calc, fadings, decorations.pathreplacing, patterns, decorations.pathmorphing, positioning}
\usepackage{tikz-cd}
\usepackage[inline]{enumitem}

\usepackage{fixmath} 
\usepackage{scalerel}
\newlength\bshft
\def\fakebold#1{\ThisStyle{\ooalign{$\SavedStyle#1$\cr%
  \kern-\bshft$\SavedStyle#1$\cr%
  \kern\bshft$\SavedStyle#1$}}}

\usetikzlibrary{positioning,shapes}
\usetikzlibrary{chains}
\usetikzlibrary{arrows,fit,decorations.pathreplacing}
\tikzstyle{every picture}+=[remember picture]
\tikzstyle{na} = [baseline=-.5ex]

\usepackage{empheq}
\usepackage{multirow}
\usepackage{booktabs}
\usepackage[american]{babel}

\usepackage[latin1]{inputenc}

\usepackage{array,booktabs}

\makeatletter
\newcommand{\vast}{\bBigg@{1}}
\newcommand{\Vast}{\bBigg@{5}}
\makeatother

\usepackage{hyperref}
\renewcommand{\arraystretch}{1.2}
\numberwithin{equation}{section}

\newcommand{\cf}{\textit{cf.}}
\newcommand{\eg}{\textit{e.g.}}

\newcommand{\ie}{\textit{i.e.}}
\newcommand{\ala}{\textit{\`a la}}

\newcommand{\ii}{\mathrm{i}}
\newcommand{\?}{\;\!}

\numberwithin{equation}{section}

\newcommand{\be}{\begin{equation}} \newcommand{\ee}{\end{equation}}
\newcommand{\bea}{\begin{equation} \begin{aligned}} \newcommand{\eea}{\end{aligned} \end{equation}}

\newcommand{\Iprod}[2]{\langle {#1}, {#2} \rangle}

\DeclareMathOperator{\sech}{sech}

\def\U{\mathrm{U}}
\def\SO{\mathrm{SO}}

\def\SU{\mathrm{SU}}

\def\PSL{\mathrm{PSL}}

\def\USp{\mathrm{USp}}

\newcommand{\rd}{\mathrm{d}}

\newcommand{\vol}{\mathrm{vol}}

\newcommand{\wb}{\overline}
\newcommand{\wt}{\widetilde}

\DeclareMathOperator{\sign}{sign}

\DeclareMathOperator{\im}{\mathbb{I}m}

\newcommand{\pd}{\partial}

\newcommand{\cA}{\mathcal{A}}
\newcommand{\cB}{\mathcal{B}}
\newcommand{\cC}{\mathcal{C}}

\newcommand{\cE}{\mathcal{E}}
\newcommand{\cF}{\mathcal{F}}

\newcommand{\cH}{\mathcal{H}}
\newcommand{\cI}{\mathcal{I}}
\newcommand{\cJ}{\mathcal{J}}
\newcommand{\cK}{\mathcal{K}}
\newcommand{\cL}{\mathcal{L}}

\newcommand{\cN}{\mathcal{N}}

\newcommand{\cR}{\mathcal{R}}
\newcommand{\cS}{\mathcal{S}}

\newcommand{\cV}{\mathcal{V}}

\newcommand{\cX}{\mathcal{X}}

\newcommand{\bR}{\mathbb{R}}

\newcommand{\fg}{\mathfrak{g}}

\newcommand{\fn}{\mathfrak{n}}

\newcommand{\fs}{\mathfrak{s}}
\newcommand{\ft}{\mathfrak{t}}

\usepackage[Symbol]{upgreek}
\usepackage{bm}

\usepackage{calligra}
\DeclareMathAlphabet{\mathcalligra}{T1}{calligra}{m}{n}

\theoremstyle{plain}

  \theoremstyle{definition}

\providecommand{\examplename}{Example}
\providecommand{\theoremname}{Theorem}

\makeatletter
\g@addto@macro\bfseries{\boldmath}
\makeatother

\makeatletter
\newcommand*{\rom}[1]{\expandafter\@slowromancap\romannumeral #1@}
\makeatother

\title{4d F(4) gauged supergravity \\ and black holes of class $\mathcal{F}$}

\author[a]{Seyed Morteza Hosseini}
\author[b,c]{and Kiril Hristov}
\affiliation[a]{Kavli IPMU (WPI), UTIAS, The University of Tokyo, Kashiwa, Chiba 277-8583, Japan}
\affiliation[b]{Faculty of Physics, Sofia University, 5 James Bourchier Blvd., Sofia 1164, Bulgaria}
\affiliation[c]{INRNE, Bulgarian Academy of Sciences, 72 Tsarigradsko Chaussee, Sofia 1784, Bulgaria}
\emailAdd{morteza.hosseini@ipmu.jp}
\emailAdd{khristov@phys.uni-sofia.bg}

\abstract{We perform a consistent reduction of 6d matter-coupled F(4) supergravity on a compact Riemann surface $\Sigma_\mathfrak{g}$ of genus $\mathfrak{g}$, at the level of the bosonic action. The result is an $\mathcal{N}=2$  gauged supergravity coupled to two vector multiplets and a single hypermultiplet. The four-dimensional model is holographically dual to the 3d superconformal field theories of class $\mathcal{F}$, describing different brane systems in massive type IIA and IIB wrapped on $\Sigma_\mathfrak{g}$. The naive reduction leads to a non-standard 4d mixed duality frame with both electric and magnetic gauge fields, as well as a massive tensor, that can be only described in the embedding tensor formalism. Upon a chain of electromagnetic dualities, we are able to determine the scalar manifolds and electric gaugings that uniquely specify the model in the standard supergravity frame. We then use the result to construct the first examples of static dyonic black holes in AdS$_6$ and perform a microscopic counting of their entropy via the 5d topologically twisted index. Finally, we show the existence of further subtruncations to the massless sector of the 4d theory, such as the Fayet-Iliopoulos gauged $T^3$ model and minimal gauged supergravity. We are in turn able to find new asymptotically AdS$_4$ solutions, providing predictions for the squashed $S^3$ partition functions and the superconformal and refined twisted indices of class $\mathcal{F}$ theories.}

\begin{document}

\setcounter{tocdepth}{2}
\maketitle

%
%

\date{Dated: \today}




\section{Introduction}
\label{sec:intro}

Five-dimensional superconformal field theories (SCFTs) are inevitably strongly coupled \cite{Chang:2018xmx}
and their string theory embeddings allow us to study the infrared (IR) phases of them.
They can be realized both in type IIA and IIB string theory.
The Seiberg theory is a $\USp(2N)$ gauge theory with $N_f$ hypermultiplets in the fundamental representation and an antisymmetric matter field, which has a 5d ultraviolet (UV) fixed point with $\SU(2)_M \times E_{N_f + 1}$ global symmetry \cite{Seiberg:1996bd}.
The theory describes the dynamics of $N$ D4-branes probing a stack of D8-branes and O8-plane, and is holographically dual to AdS$_6 \times_w S^4$ background of massive type IIA supergravity \cite{Brandhuber:1999np,Bergman:2012kr}.
A second and larger class of SCFTs can be engineered in type IIB using 5-brane webs \cite{Aharony:1997ju,Kol:1997fv,Aharony:1997bh} and it can also be generalized to include 7-branes \cite{DeWolfe:1999hj}.
The dual supergravity backgrounds \cite{DHoker:2016ujz,DHoker:2016ysh,DHoker:2017mds,DHoker:2017zwj}%
\footnote{T-duals of the massive type IIA solution~\cite{Brandhuber:1999np} have been discussed in~\cite{Lozano:2012au,Lozano:2013oma}.}
have a warped AdS$_6 \times S^2 \times_w \Sigma$ geometry, where $\Sigma$ is a Riemann surface that encodes the structure of the associated 5-brane web.

All these constructions can be realized on a compact Riemann surface $\Sigma_\fg$ of genus $\fg$ with a partial topological twist, see \cite{Bah:2018lyv} for the type IIA embedding.
Since the 5d theory has eight real supercharges with a $\SU(2)$ R-symmetry and the structure group of $\Sigma_\fg$ is $\SO(2)$, we can preserve $\cN = 2$ supersymmetry in three dimensions.
An interesting generalization is when the 5d theory has continuous flavor symmetries.
In this case, apart from the R-symmetry, we can twist the flavor symmetries by turning on various background magnetic fluxes on $\Sigma_\fg$ for these symmetries while preserving supersymmetry.
The three-dimensional SCFTs obtained in this way are called theories of class $\cF$ \cite{Bah:2018lyv}.
It is our goal here to present an effective four-dimensional $\cN = 2$ gauged supergravity model that captures the gravitational dual to the theories of class $\cF$. The model follows from a consistent truncation of six-dimensional matter-coupled F(4) gauged supergravity \cite{Romans:1985tw,Andrianopoli:2001rs,DAuria:2000xty} on $\Sigma_\fg$, at the level of the bosonic action.
We add one abelian vector multiplet that can be precisely thought of as an extra flavor symmetry of the \emph{parent} 5d theory.

The 4d F(4) gauged supergravity we obtain here may be used to construct new interesting solutions, many of which cannot be found easily in six dimensions.
The AdS$_4 \times \Sigma_\fg$ solutions of minimal F(4) supergravity were analyzed in \cite{Naka:2002jz} (see also \cite{Kim:2019fsg})
and these were generalized to the matter-coupled theory in \cite{Karndumri:2015eta,Hosseini:2018usu}.
A rotating black hole with one electric charge, two independent angular momenta and the near-horizon geometry AdS$_2 \times_w S^4$ was given in \cite{Chow:2008ip}.
The twisted AdS black hole in minimal F(4) supergravity was obtained in \cite{Suh:2018tul} and a generalization to include one independent magnetic charge, was constructed in \cite{Hosseini:2018usu,Suh:2018szn}.
In this paper, we provide the first examples of supersymmetric, asymptotically AdS$_6$, black hole solutions that carry both electric and magnetic charges,
and derive their Bekenstein-Hawking entropy from a microscopic counting of supersymmetric states in a holographically dual field theory.
New classes of Kerr-Newman-AdS (KN-AdS) black holes in 4d (that upon uplift to 6d have the exotic horizon geometry AdS$_2 \times_w S^2 \times \Sigma_\fg$)
and rotating twisted black holes are also presented whose entropy gives a prediction for the large $N$ limit of the superconformal and refined twisted indices of class $\cF$ theories, respectively.

An important role in this work is played by electromagnetic duality, which is a specific feature of four-dimensional supergravity. It is of course easy to see that the dual of a 1-form in 4d is another 1-form. The standard 4d supergravity formulation makes a certain choice to use a set of {\it electric} gauge potentials, and the equivalence of all other choices translates into the freedom to perform further duality transformations. This is known as electromagnetic duality, and the choice of electromagnetic frame is also known as a {\it duality frame}. This leads to a certain difficulty for us when reducing 6d gauge fields, since the natural reduction ansatz cannot tell us whether the resulting field is an {\it electric} or a {\it magnetic} gauge potential from the point of view of the standard supergravity formulation. 

We find that the natural duality frame from 6d perspective involves the magnetic gauge field $\tilde A_0$, whose electric counterpart we denote as $A^0$, and thus the gauge kinetic terms involve the magnetic field strength $G_0$ instead of its electric counterpart $F^0$,
\be
	A^0_\mu \leftrightarrow \tilde A_{0, \mu}\ , \qquad F^0_{\mu\nu} \leftrightarrow G_{0, \mu\nu}\ .
\ee
 In the standard frame the gauging of the internal isometries in the theory is purely electric, but from the point of view of $\tilde A_0$ coming from the reduction, the gauging is actually {\it magnetic} (using the dual gauge field $A^0$ instead of the gauge field $\tilde A_0$). It then follows that the Lagrangian can only be described using the embedding tensor formalism \cite{deWit:2002vt,deWit:2005ub} and requires additional couplings to a tensor field. Due to the existence of a two-form field $B$ in the six-dimensional gauged supergravity, we naturally obtain such a two-form field in four dimensions. It turns out that the universal hypermultiplet (UHM) can be actually dualized to a tensor-scalar multiplet as explained in \cite{Theis:2003jj} and \cite{DallAgata:2003sjo,DAuria:2004yjt}. Schematically, we have 
\be
	\sigma \leftrightarrow B_{\mu\nu}\ ,
\ee
where $\sigma$ is one of the hypermultiplet scalars, and the dualization is ``carried out'' by the gauge field $A^0$ in a precise way explained in the main text. It is then interesting to note that the Higgs mechanism where the original massless scalar $\sigma$ is eaten up by a vector field, in this new duality frame is replaced by a Stueckelberg mechanism where the same vector field in question instead gets eaten up by the two-form $B$ that becomes massive itself. 

When brought to the standard duality frame, we find an $\cN=2$ supergravity coupled to two vector multiplets and the UHM with a particular electric gauging of two of the internal hypermultiplet isometries. The outcome is that at the AdS$_4$ vacuum one of the vectors becomes massive and together with a complex vector multiplet scalar and the hyperscalars forms a massive vector multiplet.  The form of the prepotential defining the remaining massless scalar manifold and depending on the magnetic fluxes through the Riemann surface then agrees with the one predicted in \cite{Hosseini:2018usu} from a direct six-dimensional computation. Just like any matter-coupled theory, the resulting model admits a further subtruncation to minimal gauged supergravity. For some specific values of the magnetic fluxes the massive modes can also be truncated to a larger sector, the so-called $T^3$ model, with a single massless vector multiplet and Fayet-Iliopoulos (FI) gauging. 

For the reader's convenience, we summarize our main results regarding the 4d supergravity model in the next subsection.
The reader not interested in the details of the dimensional reduction and further subtruncations, can read it and then safely jump to sections \ref{sec:universal} and \ref{sec:BlackHoles}.

The remainder of this paper is organized as follows. In section \ref{sec:6donSigma} we give an overview of 6d matter-coupled F(4) supergravity and define our reduction ansatz on $\Sigma_\fg$. We next derive the effective 4d action.
In section \ref{sec:4d} we formulate the 4d truncated theory in a canonical supergravity language, discussing how one can switch between different duality frames.
In section \ref{sec:trunc} we discuss the possibility for further consistent truncations to subsectors of the original 4d supergravity, identifying explicitly two such cases. In section \ref{sec:universal} we discuss asymptotically AdS$_4$ solutions in the minimal subtruncation related to a universal subsector of the class $\cF$ theories. In section \ref{sec:BlackHoles} we write down general classes of supersymmetric black hole solutions, summarizing previously known cases and extending them to new solutions with electric charges and rotation. We then discuss their corresponding entropy functions from gluing gravitational blocks in section \ref{sec:gluing}, and in the process provide new holographic predictions for two different classes of supersymmetric indices.  Finally, in section \ref{sec:conclusions}, we conclude with a further field theory discussion and a list of open problems. 

\subsection{Main results}
\label{subsec:results}

In the standard duality frame we find an ${\cal N}=2$ supergravity coupled to two vector multiplets with scalar manifold $[\SU(1,1)/ \U(1)]^2$, and one hypermultiplet, known as the universal hypermultiplet (UHM), with scalar manifold $\SU(2,1)/\U(2)$. The bosonic field content is given by a metric, three $\U(1)$ gauge fields $A^{0,1,2}$, two complex scalars with an underlying special K\"ahler manifold defined by the prepotential 
\be
\label{F:incomplete}
	F =- \ii X^0 \sqrt{X^1 X^2} \, ,
\ee
and four real scalars $\{\phi,\sigma, \zeta, \tilde{\zeta} \}$ spanning the UHM  with a well-known quaternionic metric. The three gauge fields are used for the gauging of two of the isometries of the UHM manifold. The first isometry is compact and is gauged by a particular combination of the electric gauge fields given by the following Killing vector in symplectic notation,
\be
	k^{\U(1)} = \{0; 0, 3 m , 3 m \} \, ,
\ee
while the second gauged isometry is non-compact and is defined by the Killing vector
\be
	k^\mathbb{R} = \left\{ 0 ; - 4m , s^2 , s^1 \right\} .
\ee
The parameter $m$ denotes the gauge coupling constant that controls the radius of the internal manifold from a ten-dimensional perspective,
and $s^{1,2}$ are the magnetic fluxes through the Riemann surface with a constant curvature $\kappa$, satisfying the twisting condition
\be
	s^1 + s^2 = - \frac{\kappa}{3m}\, .
\ee
The above data is enough to uniquely specify the Lagrangian of the theory, with all explicit details spelled out in the main sections of this paper.

Due to the non-compact gauging, a spontaneous symmetry breaking (but fully supersymmetry preserving) Higgs mechanism takes place for many background solutions, including the maximally supersymmetric AdS$_4$ vacuum. The particular combination of the vector fields as defined by $k^\mathbb{R}$ becomes massive by eating up the scalar $\sigma$, and together with other five massive scalars%
\footnote{The five massive scalars are the three remaining hyperscalars $\phi, \zeta, \tilde \zeta$ together with one of the two complex scalars. In our dimensional reduction we actually set to zero the two charged 6d scalars proportional to the 4d $\zeta, \tilde \zeta$. It is easy to see that this is a consistent subtruncation of the equations of motion and will not prevent us from properly determining the 4d supergravity structure.} forms a massive vector multiplet \cite{Hristov:2010eu}. The effective prepotential for the massless field at the BPS Higgsed locus is then given by
\be
	F_{\text{eff}} = - \frac{\ii}{4 m} \left( s^2 X^1 + s^1 X^2 \right) \sqrt{X^1 X^2} \, ,
\ee
in agreement with the expected supergravity action for the vacuum AdS$_4$ solution \cite[\cf\,(4.8)]{Hosseini:2018usu}.
Note that this is not a supersymmetric truncation in general, and the effective prepotential is a useful description of the massless sector strictly at the considered background.

In the special case when one of the parameters $s^{1,2}$ vanishes, one is actually able to perform a consistent supersymmetric subtruncation to the purely massless sector already at the level of the theory, such that the effective prepotential above truly becomes the prepotential of the theory. This special limit of parameters leads to the so-called $T^3$ model with a scalar manifold  $\SU(1,1)/ \U(1)$ with constant FI gauge parameters (as the hypermultiplet has been truncated out), and the prepototential
\be
 F_{T^3} = - \frac{\ii}{12 m^2} \sqrt{X^1 (X^2)^3} \, .
\ee
The same model can alternatively be obtained from a consistent truncation of maximal $\cN = 8$ supergravity, where many BPS solutions have already been constructed. We can then directly borrow these results and embed them in 10d massive type IIA and type IIB supergravity, producing a variety of new bulk solutions and holographic predictions.

Finally, let us also note that already the general model, without any restrictions on $s^{1,2}$, admits a consistent truncation to minimal $\cN=2$ gauged supergravity. This is of course guaranteed by the general structure of supergravity, but also allows for further holographic predictions if one restricts to the universal sector of the dual field theory, as in \cite{Azzurli:2017kxo,Bobev:2017uzs}.

\section[6d F(4) supergravity on \texorpdfstring{$\Sigma_\fg$}{Sigma[g]}]{6d F(4) supergravity on $\Sigma_\fg$}
\label{sec:6donSigma}

\subsection{Matter coupled F(4) gauged supergravity}
\label{subsec:6d}

We first briefly review six-dimensional F(4) gauged supergravity \cite{Romans:1985tw} coupled to $n_{\text{V}}$ vector multiplets \cite{Andrianopoli:2001rs,DAuria:2000xty}.
The field content of the gauged supergravity theory is as follows.
The bosonic part of the gravity multiplet consists of the metric $g_{\mu\nu}$, four gauge fields $\cA^\alpha_\mu$ corresponding to the symmetry group $\U(1) \times \SU(2)_R$ where the latter factor is the R-symmetry, an antisymmetric tensor field $B_{\mu\nu}$, and the dilaton $\lambda$.
The fermionic components are two gravitini $\psi_\mu^A$, and two spin one-half fermions $\chi^A$ in the fundamental representation of $\SU(2)_R$.
It is useful to split the index $\alpha=(0,r)$ where $r=1,2,3$ is an index in the adjoint representation of $\SU(2)_R$.
The gravity multiplet can be coupled to $n_{\text{V}}$ vector multiplets, which are labeled by an index $I = 1, \ldots, n_{\text{V}}$.
Each vector multiplet contains one vector field $\cA_\mu$, four scalars $\phi_\alpha$ and  two gaugini $\lambda_A$.
The $4 n_\text{V}$ scalar fields parameterize the coset space
$
 \frac{\SO(4,n_\text{V})}{\SO(4)\times \SO(n_\text{V})} \, .
$
It is convenient to introduce a new index $\Lambda=(\alpha,I)$ and encode the scalar fields into a coset representative $L^\Lambda_{\phantom{\Lambda}\Sigma} \in \SO(4,n_\text{V})$.

The bosonic action, following the conventions of \cite{Andrianopoli:2001rs}, reads
\be
	S_{6\rd} = \frac1{G_{\text{N}}^{6\rd}}\, \int\, \rd^6 x\, \sqrt{-g}\, \cL_{6\rd}\ ,
\ee
with
\bea
 \label{Lagrangian6d}
 \cL_{6\rd} = & - \frac14R -\frac18 e^{-2 \lambda}{\cal N}_{\Lambda\Sigma} \hat \cF_{\mu\nu}^\Lambda \hat \cF^{\Sigma\mu\nu} + \frac{3}{64} e^{4 \lambda} H_{\mu\nu\rho}H^{\mu\nu\rho} + \partial^\mu \lambda \partial_\mu \lambda -\frac14 P^{I\alpha\mu} P_{I\alpha\mu} \\
 & - \frac{1}{64} \varepsilon^{\mu\nu\rho\sigma\lambda\tau} B_{\mu\nu} \left( \eta_{\Lambda\Sigma} \hat \cF_{\rho\sigma}^\Lambda \hat \cF_{\lambda\tau}^\Sigma + m B_{\rho\sigma}  \hat \cF_{\lambda\tau}^0 +\frac13 m^2 B_{\rho\sigma}  B_{\lambda\tau} \right) + V_{6\rd} (\lambda, \phi_\alpha)  \, ,
\eea 
where
\bea
 & \hat \cF_{\mu\nu}^\Lambda = \cF_{\mu\nu}^\Lambda - m \delta^{\Lambda 0} B_{\mu\nu} \, , \\
 & {\cal N}_{\Lambda\Sigma} = L_\Lambda^{\phantom{\Lambda}\alpha} ( L^{-1})_{\alpha\Sigma} - L_\Lambda^{\phantom{\Lambda}I} ( L^{-1})_{I\Sigma} \, , \\
 & P^I_\alpha = (L^{-1})^I_{\phantom{I}\Lambda} \left( \rd L^\Lambda_{\phantom{\Lambda}\alpha} - f_{\Gamma\phantom{\Lambda}\Pi}^{\phantom{\Pi}\Lambda} \cA^\Gamma L^\Pi_{\phantom{\Pi}\alpha} \right) ,
\eea
with $f^\Lambda_{\phantom{\Lambda}\Pi\Gamma}$ the structure constants of the gauge group $\SU(2)_R\times G$, and the $\SO(4,n_\text{V})$ invariant metric $\eta_{\Lambda \Sigma} = {\rm diag}\{ 1,1,1,1,-1,\dots, -1\}$.
The gravitini are charged under the $\SU(2)_R$ with a charge $g$, while $m$ corresponds directly to the mass parameter of massive type IIA supergravity \cite{Cvetic:1999un}.

The particular theory we consider contains one vector multiplet, $n_\text{V}=1$, corresponding to the $\U(1) \subset \SU(2)$ mesonic symmetry of the holographic dual five-dimensional SCFT.
We consistently set to zero all vector fields except $\cA^0_\mu$, $\cA^{r=3}_\mu$ in $\SU(2)_R$ (thus breaking $\SU(2)_R$ to a $\U(1)_R$), and $\cA^{I=1}_\mu$.
Further, we require that the scalar fields $\phi_{\alpha}$ in the vector multiplet to be neutral under $\cA^{r=3}_\mu$ and this leaves us with $\phi_0$ and $\phi_3$.%
\footnote{Setting to zero the two scalars $\phi_1$ and $\phi_2$, charged under the remaining $\cA^3_\mu$ of the original $\SU(2)$, is strictly speaking a further truncation that we are allowed to perform only at the level of the bosonic equations of motion. We will see during the reduction that this has its precise correspondence to the resulting four-dimensional supergravity model where the two charged scalars in the hypermultiplet, standardly denoted by $\zeta$ and $\tilde \zeta$, have been also set to zero in a consistent way.}
A convenient parameterization of the scalar coset is given by \cite{Karndumri:2015eta,Gutperle:2017nwo,Gutperle:2018axv}
\be
 L^\Lambda_{\phantom{\Lambda} \Sigma} =
 \begin{pmatrix}
  \cosh \phi_0 & 0 & 0 & \sinh \phi_0 \sinh \phi_3 & \sinh \phi_0 \cosh \phi_3 \\ 0 & 1 & 0 & 0 & 0 \\ 0 & 0 & 1 & 0 & 0 \\ 0 & 0 & 0 & \cosh \phi_3 & \sinh \phi_3 \\ \sinh \phi_0 & 0 & 0 & \cosh \phi_0 \sinh \phi_3 & \cosh \phi_0 \cosh \phi_3
 \end{pmatrix} \, .
\ee
The kinetic terms for the vectors can then be written as
{\small
\be {\cal N}_{\Lambda\Sigma} =
 \begin{pmatrix}
  \cosh^2 \phi_0 + \cosh (2 \phi_3) \sinh^2 \phi_0 & 0 & 0 & \sinh \phi_0 \sinh (2 \phi_3) &  -\cosh^2 \phi_3 \sinh (2 \phi_0 )\\ 0 & 1 & 0 & 0 & 0 \\ 0 & 0 & 1 & 0 & 0 \\  \sinh \phi_0 \sinh (2 \phi_3) & 0 & 0 & \cosh (2 \phi_3 )& -\cosh \phi_0 \sinh( 2 \phi_3) \\ -\cosh^2 \phi_3 \sinh (2 \phi_0) & 0 & 0 & -\cosh \phi_0 \sinh( 2 \phi_3) & \cosh^2 \phi_0 \cosh( 2 \phi_3) + \sinh^2 \phi_0
 \end{pmatrix} \, .
\ee 
}
Finally, the scalar potential for the fields we have chosen to keep reads
\be
	V_{6\rd} =  g^2 e^{2 \lambda} - m^2 e^{-6 \lambda} ( \cosh^2 \phi_0 + \sinh^2 \phi_0 \cosh (2 \phi_3) )  \\ 
+ 4 m g\? e^{-2 \lambda} \cosh \phi_0 \cosh \phi_3 \, .
\ee
The potential admits an AdS$_6$ background for $g=3m$, as the pure F(4) gauged supergravity \cite{Romans:1985tw,Andrianopoli:2001rs,DAuria:2000xty}. For later convenience we also define the following linear combinations of the vector fields
\bea
	& A^{1,2}_\mu \equiv \cA^3_\mu \pm \cA^4_\mu \quad \Rightarrow \quad \cA^{3,4} = \frac12 (A^1_\mu \pm A^2_\mu) \, , \\
	& A_\mu^{0} \equiv \cA_\mu^0\, .
\eea

We believe that this theory as presented above is a consistent truncation of massive type IIA supergravity on the warped background AdS$_6 \times S^4$. Ample evidence for this was presented in \cite{Hosseini:2018usu}: on one hand the six-dimensional AdS$_4 \times \Sigma_\fg$ solutions were shown to exactly agree with the direct ten-dimensional solutions of \cite{Bah:2018lyv}; on the other hand the corresponding on-shell action and/or macroscopic entropy of different supersymmetric asymptotically AdS$_6$ solutions were successfully matched to the appropriate partition functions of the five-dimensional Seiberg theory computed in \cite{Hosseini:2018uzp} and \cite{Crichigno:2018adf}. Additionally, \cite{Malek:2019ucd} showed that this theory arises as a consistent truncation from type IIB supergravity on a general class of manifolds, including the abelian T-dual of the type IIA background in question.

\subsection{The truncation ansatz}
\label{subsec:reduction}
We now turn to construct the consistent bosonic truncation of the 6d F(4) gauged supergravity coupled to a single vector multiplet on a smooth Riemann surface of genus $\fg$, down to 4d $\cN = 2$ gauged supergravity. The ans\"atze for the reduction of the various bosonic fields are quite natural and follow similar consistent reductions on Riemann surfaces in other settings, see for example \cite{Szepietowski:2012tb,Cheung:2019pge}. 

We choose the truncation ansatz for the metric such that we preserve four-dimensional Einstein frame,
\be
	{\rm d} s_6^2 = {\rm e}^{-2 C} \rd s_4^2 - {\rm e}^{2 C} \rd s_{\Sigma_\fg}^2 \, ,
\ee
where $C$ is the additional Kaluza-Klein (KK) scalar.
We choose a constant curvature metric on the Riemann surface%
\footnote{We omit the more detailed discussion of the torus, $\fg = 1$, as it requires a slightly more careful treatment of quantities such as $(1-\fg)$ and $\kappa$. It is however clear that the toroidal case does not present a conceptual problem and can easily be incorporated in the analysis, see e.g.\ \cite{Hosseini:2018usu}. We leave this as a simple exercise to the interested reader.}
\be
 \rd s^2_{\Sigma_\fg} = e^{2h(x,y)} (\rd x^2 + \rd y^2) \, , \quad \text{ with } \quad
 e^{2 h (x,y)} =
 \Bigg\{\begin{array}{lr}
        \frac{4}{(1 + x^2 + y^2)^2} \, , & \text{ for } S^2 \\
        \frac1{y^2} \, , & \text{ for } H^2
        \end{array} \, .
\ee
In our convention, we have $R_{\Sigma_\fg} = 2 \kappa$ with $\kappa = +1$ for $S^2$ and $\kappa = -1$ for $H^2$.
The coordinates $(x,y)$ take values in $\bR^2$ for $S^2$ and in $\bR \times \bR_{>0}$ for $H^2$.
In the latter case the upper half-plane is quotiented by a suitable Fuchsian subgroup $\Gamma \subset \PSL(2,\bR)$ to obtain a compact Riemann surface $\Sigma_{\fg > 1}$.
The volume form 
integrates to
\be
 \label{volSigma}
 \vol (\Sigma_\fg) = \int \vol_{\Sigma_\fg} = \int e^{2h(x,y)} \rd x \wedge \rd y = 4 \pi | \fg - 1 | \, .
\ee

The scalars are reduced in a straightforward fashion,
\be
	\lambda_{6\rd} = \lambda_{4\rd} \, , \qquad \phi_{0, 6\rd} = \phi_{0, 4\rd} \, , \qquad \phi_{3, 6\rd} = \phi_{3, 4\rd} \, .
\ee
The gauge fields and the corresponding field strengths are reduced as follows
\be
	A^i_{6\rd} = 2\? A_{4\rd}^i + 2\? \kappa\? s^i \omega_\Sigma\, , \qquad F^i_{6\rd} = 2\? F^i_{4\rd} + 2\? s^i \vol_{\Sigma_\fg} \, , \quad i = 1, 2 \, ,
\ee
with $\omega_{\Sigma_\fg}$ being the spin-connection on the Riemann surface. Supersymmetry further dictates that \cite{Hosseini:2018usu}
\be
	s^1 + s^2 = - \frac{\kappa}{3m}\, .
\ee
The vector field $\cA^0$ is reduced as
\be
 \label{ansatz:eTm}
	A^0_{6\rd} = 2\? \tilde A_{0, 4\rd}\, , \qquad F^0_{6\rd} = 2\? G_{0, 4\rd}\, .
\ee
Note that in \eqref{ansatz:eTm} we have intentionally lowered the index of the four-dimensional fields and have used different letters to denote them,
anticipating the major complication of the resulting non-standard duality frame for the reduced gauge field.
We explain how this comes about carefully in the next section.

Finally, we reduce the antisymmetric tensor field as
\be
	B_{6\rd} = - 2 B_{4\rd}+ 2\? b_{4\rd}\, \vol_{\Sigma_\fg} \ , \qquad H_{6\rd} = - 2 H_{4\rd} + \frac{2}{3} \rd b_{4\rd} \wedge \vol_{\Sigma_\fg} \ .
\ee

\subsection{Effective 4d theory}
\label{subsec6dto4d}

Using the above ansatz for the reduction, we can write out everything in terms of the four-dimensional fields and explicitly perform the integrals in the action over the Riemann surface coordinates. Dropping for clarity the superfluous index specifying that the fields are now in four dimensions, we arrive at
\be
	S^{\text{eff}}_{4\rd} = \frac1{G_{\text{N}}^{4\rd}}\, \int\, \rd^4 x\, \sqrt{-g}\, {\cal L}^{\text{eff}}_{4\rd}\, , \qquad \frac1{G_{\text{N}}^{4\rd}} = \frac{\vol (\Sigma_\fg)}{G_{\text{N}}^{6\rd}}\, ,
\ee
\bea
\label{eq:reducedLag}
\begin{split}
	{\cal L}^{\text{eff}}_{4\rd} &= -\frac14 R + (\partial_\mu \lambda)^2 +\frac14\cosh^2 \phi_3 \? (\partial_\mu \phi_0)^2 + \frac14 (\partial_\mu \phi_3)^2 + (\partial_\mu C)^2 \\
& - \frac18\, e^{2 C -2 \lambda}\Big[ 4 ( \cosh^2 \phi_0 + \cosh (2 \phi_3) \sinh^2 \phi_0) (\hat G_{0, \mu \nu})^2  + \cosh (2 \phi_3) (F^1_{\mu \nu}+F^2_{\mu \nu})^2 \\
&+ (\cosh^2 \phi_0 \cosh (2\phi_3) + \sinh^2 \phi_0 ) (F^1_{\mu \nu}-F^2_{\mu \nu})^2 - 2 \cosh \phi_0 \sinh (2\phi_3) ((F^1_{\mu \nu})^2 - (F^2_{\mu \nu})^2) \\
& + 4 \sinh \phi_0 \sinh(2 \phi_3) \hat G_0^{\mu \nu} (F^1_{\mu \nu}+F^2_{\mu \nu}) - 4 \sinh (2 \phi_0) \cosh^2 \phi_3 \hat G_0^{\mu \nu} (F^1_{\mu \nu} - F^2_{\mu \nu})  \Big]\\
& + \frac{1}{8}\? \varepsilon^{\mu\nu\rho\sigma} B_{\mu\nu} \left(s^2 F_{\rho\sigma}^1 + s^1 F_{\rho\sigma}^2 \right)
- \frac{b}8 \varepsilon^{\mu\nu\rho\sigma} \left( \hat G_{0, \mu\nu} \hat G_{0, \rho\sigma} + F^1_{\mu\nu} F_{\rho\sigma}^2 \right) \\
& +\frac3{16}\, e^{4 (\lambda + C)} ( H_{\mu\nu\rho})^2 + \frac{1}{32} e^{4 (\lambda - C)} (\partial_\mu b)^2 + V^{\text{eff}}_{4\rd}(C,\lambda, \phi_0, \phi_3) \, ,
\end{split}
\eea
with
$
 \hat G_0 = G_0 + m B \, ,
$
and the scalar potential
\bea
 \label{Veff}
\begin{split}
	V^{\text{eff}}_{4\rd} &= \frac{\kappa}2\, e^{-4 C} +m^2 e^{-2 C} \Big[ 9 e^{2 \lambda} - e^{-6 \lambda} ( \cosh^2 \phi_0 + \sinh^2 \phi_0 \cosh (2 \phi_3) )  \\ 
&+ 12 e^{-2\lambda} \cosh \phi_0 \cosh \phi_3 \Big] - \frac{e^{-2\lambda-6 C}}4\, \Big[ \frac{\cosh (2 \phi_3)}{4} (s^1+s^2)^2 \\
&+ \frac{\cosh^2 \phi_0 \cosh (2\phi_3) + \sinh^2 \phi_0}{4} (s^1-s^2)^2 - \frac{\cosh \phi_0 \sinh (2\phi_3) }{2} ((s^1)^2 - (s^2)^2)  \Big] \\
&  - \frac{m}4\, e^{-2 \lambda - 6 C}\Big[ m\? b^2 ( \cosh^2 \phi_0 + \cosh (2 \phi_3) \sinh^2 \phi_0) \\ 
& - b \sinh \phi_0 \sinh(2 \phi_3) (s^1 + s^2)  + b \sinh (2 \phi_0) \cosh^2 \phi_3\, (s^1 - s^2)  \Big] \, .
\end{split}
\eea

\section[4d \texorpdfstring{$\cN = 2$}{N=2} supergravity structure]{4d $\cN = 2$ supergravity structure}
\label{sec:4d}

\subsection{The duality frame conundrum}
\label{subsec:duality}

Having in mind the form of the reduced Lagrangian \eqref{eq:reducedLag}, it is at first sight rather mysterious how to fit it in a standard 4d ${\cal N}=2$ supergravity language. The resolution comes by realizing that the gauge fields coming from 6d might be either electric or magnetic from the 4d point of view. In fact they turn out to be mixed, such that we can either choose $\tilde A_0$ to be magnetic and $A^{1,2}$ electric, or the exact opposite arrangement. For convenience we choose the former option (as anticipated in the choice of index position), such that in the duality frame we present the final Lagrangian we have purely electric gaugings for both isometries that are gauged. 

Let us for the moment keep full symplectic covariance of the Lagrangian with a number of electric (index $U, V$) and magnetic (index $I, J$) mutually exclusive field strengths, and we allow for general magnetic gaugings as prescribed by \cite{deWit:2005ub} that introduced the embedding tensor formalism. In particular, let us first consider only the vector-tensor part of the Lagrangian to illustrate more concisely our main point:%
\footnote{Note that we follow the conventions imposed to us from the starting six-dimensional theory \cite{Andrianopoli:2001rs}. In particular this means we have the less often used $F_{\mu\nu} =\frac12 ( \partial_\mu A_\nu - \partial_{\nu} A_\mu)$ and subsequently need to adapt the four-dimensional theory of \cite{deWit:2005ub} in the same conventions.}
\begin{eqnarray}
\label{eq:new-Ltotal}
\begin{split}
  {\cal L}_\text{V-T} &=
  \frac12 \Big[ \hat{\cal I}^{IJ}\,\hat{G}_{\mu\nu I} \,
   \hat{G}^{\mu\nu}{}_J + \hat{\cal
   I}_{UV}\,\hat{F}_{\mu\nu}{}^U\, 
   \hat{F}^{\mu\nu V} + 2\,\hat{\cal I}^I{}_U
   \,\hat{G}_{\mu\nu I}\, \hat{F}^{\mu\nu U} \Big]    
 \\[.9ex]
  &{}
  + \frac14 \varepsilon^{\mu\nu\rho\sigma} 
  \Big[ \hat{\cal R}^{IJ}\,\hat{G}_{\mu\nu I} \,
   \hat{G}_{\rho\sigma J} + \hat{\cal
   R}_{UV}\,\hat{F}_{\mu\nu}{}^U\, \hat{F}_{\rho\sigma}{}^V 
   + 2\,\hat{\cal R}^I{}_U    \,\hat{G}_{\mu\nu I} \,
   \hat{F}_{\rho\sigma}{}^U \Big]    \\[.9ex]
& -\frac18\, \varepsilon^{\mu\nu\rho\sigma}\,
\Theta^{U \alpha}\,B_{\mu\nu\,\alpha} \,
\Big(
\partial_{\rho} \tilde A_{\sigma\, U}
-\frac18\Theta_U{}^{\beta}B_{\rho\sigma\,\beta} \Big)  \\[.9ex]
&+\frac18\, \varepsilon^{\mu\nu\rho\sigma}\,
\Theta_I{}^{\alpha}\,B_{\mu\nu\,\alpha} \,
\Big(
\partial_{\rho} A^I_{\sigma} 
-\frac18\Theta^{I \beta}B_{\rho\sigma\,\beta} \Big) \, ,
\end{split}
\end{eqnarray}
where
\be
	\hat G_{I, \mu\nu} :=  \partial_{[ \mu} \tilde A_{I, \nu]} - \frac14\, \Theta_I{}^\alpha B_{\alpha, \mu\nu}\ , \qquad \qquad  \hat F^U_{\mu\nu} :=\partial_{[\mu} A^U_{\nu ]} + \frac14\, \Theta^{U \alpha} B_{\alpha, \mu\nu}\ ,
\ee
and the matrices $\hat \cI$ and $\hat \cR$ can be produced by symplectically rotating the imaginary and real parts of the standard period matrix, which we will denote as (unhatted) ${\cal I}$ and ${\cal R}$.
We show explicitly how to perform this symplectic rotation in the next subsection. 

The embedding tensor $\Theta$ obeys a number of identities that render the theory consistent. In particular, notice that we have in principle used all electric and all magnetic gauge fields in the above Lagrangian, the electric $A^U$ corresponding to the electric field strength $F^U$ and their magnetic duals $\tilde A_U$, together with the \emph{electric} $\tilde A_I$ corresponding to the magnetic $G_I$, and their corresponding \emph{magnetic} duals $A^I$.%
\footnote{We apologize to the perplexed reader, but unfortunately in the mixed duality frame the terminology can often lead to confusion. This is because from the point of view of the magnetic field strengths $G_I$ we can still have magnetic gauging that in the standard duality frame with purely electric field strengths is electric. We chose to write the non-standard nomenclature in italics, to emphasize that this is the opposite of what one sees in the standard purely electric frame.}
The embedding tensor identities guarantee that exactly half of the gauge fields carry the fundamental degrees of freedom. It is also evident that the tensor fields appear in the definition of the composite $\hat F$ or $\hat G$ whenever there is a \emph{magnetic} gauging from the point of view of the field strength in question, and the corresponding magnetic gauge field enters in the Chern-Simons terms with the tensors. 

The abstract form of the action might easily lead to various confusions about electromagnetic duality, so we  prefer to directly discuss the example of interest to us. We only consider a single tensor field, thus allowing us to erase the index $\alpha$ altogether, and look at a case with a total of three vectors such that the first one (index $I = 0$) is magnetic and the other two electric ($U=1,2$). We further choose (anticipating the match we will make with the reduced 6d theory) the nonvanishing components of the embedding tensor to be
\be
\label{eq:embtensor}
	\Theta_0 = - 4 m\ , \qquad \Theta_1 = s^2 \, , \qquad \Theta_2 = s^1 \, .
\ee
From the standard point of view this just corresponds to a purely electric gauging, and the theory can very easily and efficiently be written in the conventions of \cite{Andrianopoli:1996cm}. However, in the duality frame we are forced to work from the 6d reduction, the first component above specifies a \emph{magnetic} gauging from the point of view of the magnetic field strength $G_0$. We therefore have
\be
	\hat G_0 = G_0 + m B\ , \qquad \qquad \hat F^{1,2} = F^{1,2}\ , 
\ee
and the resulting vector-tensor interaction term reads
\be
  {\cal L}_\text{V-T} \ni - \frac{m}2 \varepsilon^{\mu\nu\rho\sigma} B_{\mu\nu}  \partial_{\rho} A^0_{\sigma} \, .
\ee
This is however \emph{not} the final form of the Lagrangian that is ready for a comparison with the 6d reduction. Notice in particular the appearance of the gauge field $A^0$, which does not come out of the 6d reduction ansatz. In order to get to the correct duality frame we need to integrate it out using its equation of motion, but only after we have included the correct coupling to the UHM. 

To do this, let us briefly look at the UHM, which is a realization of the coset space $\SU(2,1) / \U(2)$. The metric, written in terms of real coordinates $\{\phi,\sigma, \zeta, \tilde{\zeta} \}$, is  given by
\begingroup
\renewcommand*{\arraystretch}{1.2}
\begin{equation}
h_{u v} = 
\begin{pmatrix}
1 & 0 & 0 & 0 \\
0 & \frac{1}{4} e^{4\phi} & - \frac{1}{8} e^{4\phi} \tilde{\zeta} & \frac{1}{8} e^{4\phi} \zeta \\
0 & - \frac{1}{8} e^{4\phi} \tilde{\zeta} & \frac{1}{4} e^{2\phi}(1 + \frac{1}{4} e^{2\phi} \tilde{\zeta}^2) & -\frac{1}{16} e^{4\phi} \zeta \tilde{\zeta} \\
0 &  \frac{1}{8} e^{4\phi} \zeta & -\frac{1}{16} e^{4\phi} \zeta \tilde{\zeta} &  \frac{1}{4} e^{2\phi}(1 + \frac{1}{4} e^{2\phi} \zeta^2)
\end{pmatrix}\, .
\end{equation}
\endgroup
The isometry group $\SU(2,1)$ has eight generators; two of these are used for gauging in the model we derive here, generating the group $\mathbb{R} \times \U(1)$. The corresponding Killing vectors read
\begin{align}
	k^{\mathbb{R}} = \partial_{\sigma}\, , \qquad k^{\U(1)} = -\tilde{\zeta} \partial_{\zeta} + \zeta \partial_{\tilde{\zeta}}\, .
\end{align}
These two isometries are gauged by a particular linear combination of the vector fields in the theory. We define Killing vectors with a symplectic index corresponding to each of the full set of electric and magnetic gauge fields at our disposal. The moment maps associated to these two Killing vectors are given by
\begin{align}
	P^{\mathbb{R}} = \begin{pmatrix} 0, & 0, &  -\frac{1}{2} e^{2\phi} \end{pmatrix} \, , \qquad 
P^{\U(1)} = \begin{pmatrix} \tilde{\zeta} e^{\phi}, & - \zeta e^{\phi}, &  1 - \frac{1}{4} (\zeta^2 + \tilde{\zeta}^2) e^{2\phi}\end{pmatrix} \, . 
\end{align}
In the explicit reduction from 6d we have only kept half of the hypermultiplet scalars, which amounts to the further consistent truncation
\be
	\zeta = \tilde \zeta = 0\ .
\ee
This means the hypermultiplet metric takes the simple diagonal form
\be
\label{hyper:metric}
	h_{u v} = \left( \begin{array}{cc} 1 & 0 \\ 0 & \frac14 \? e^{4 \phi} \\
\end{array} \right) ,
\ee
and the Killing vectors associated with the $\U(1)$ isometry completely vanish. There is however a residual trace of the $\U(1)$ gauging in the form of the moment maps
\be
	P^{\U(1)} = ( 0 , 0 , 1) \, . 
\ee
These moment maps show up in the expression for the scalar potential and therefore we will be able to infer their existence even in the truncation $\zeta = \tilde \zeta = 0$. Coming back to the remaining part of the hypermultiplet, the embedding tensor \eqref{eq:embtensor} uniquely determines how the scalars $\phi$ and $\sigma$ are coupled to the gauge fields in the Lagrangian as it specifies the correct linear combination of gauge fields used for gauging the non-compact isometry,%
\footnote{Note that strictly speaking we should have already specified that the components of the embedding tensor in \eqref{eq:embtensor} are only corresponding to the non-compact isometry. The compact isometry defines its own embedding tensor components that we have so far suppressed as they do not lead to extra terms in the Lagrangian. We instead specify the resulting moment maps explicitly when discussing the scalar potential.}
\be
	k^\bR = \left\{ 0 ; - 4m , s^2 , s^1 \right\} \, ,
\ee
such that the hypermultiplet scalar kinetic terms are given by
\be
	{\cal L}_\text{H} =  \frac12\? h_{uv} \nabla_\mu q^u \nabla^\mu q^u =  \frac12\? (\partial_\mu \phi)^2 + \frac18\? e^{4\phi} \left(\partial_\mu \sigma - ( 4 m A^0_\mu - s^2 A^1_\mu - s^1 A^2_\mu)\right)^2\, .
\ee 
Now we can put together \emph{all} terms in the Lagrangian featuring the gauge field $A^0$, which in the model we consider is the only one that has no kinetic term (as opposed to the gauge fields $\tilde A_0, A^{1,2}$ that make up the field strengths $G_0, F^{1,2}$),
\be
	{\cal L}_{A^0} = - \frac{m}2\? \varepsilon^{\mu\nu\rho\sigma}\? B_{\mu\nu} \?  \partial_{\rho} A^0_{\sigma} + \frac18\? e^{4\phi} \left(\partial_\mu \sigma - ( 4 m A^0_\mu - s^2 A^1_\mu - s^1 A^2_\mu) \right)^2 \, .
\ee
The equation of motion for $A^0$ therefore fixes
\be
\label{eq:Btosigma}
	\frac12\, \varepsilon_\mu{}^{\nu\rho\sigma}\, \partial _\nu B_{\rho\sigma} =  e^{4 \phi}\, \nabla_\mu \sigma\, ,
\ee
giving a duality relation between the, so far, auxiliary tensor field $B$ and the covariant derivative of the hyperscalar $\sigma$, as noticed in \cite{Guarino:2017pkw}. This means that upon integrating out $A^0$, we generate a kinetic term for the tensor field $B$  and new Chern-Simons couplings between the tensor and the electric gauge fields $A^{1,2}$ at the expense of the scalar $\sigma$ that disappears,
\be
	{\cal L}_{A^0} = \frac{3\, e^{-4 \phi}}{16}\? H_{\mu\nu\rho} H^{\mu\nu\rho}  + \frac{1}8\? \varepsilon^{\mu\nu\rho\sigma} B_{\mu\nu} \left( s^2  \partial_{\rho} A^1_{\sigma} + s^1 \partial_{\rho} A^2_{\sigma} \right) ,
\ee
where
\be
	H_{\mu\nu\rho} = \partial_{[ \mu} B_{\nu\rho]} = \frac13\, (\partial_\mu B_{\nu\rho} - \partial_\nu B_{\mu\rho} + \partial_\rho B_{\mu\nu})\, .
\ee
If we now put all components of the 4d supergravity Lagrangian together, we arrive at a rather unusual duality frame which more closely resembles the earlier constructions of \cite{Theis:2003jj,DallAgata:2003sjo,DAuria:2004yjt},
\begin{eqnarray}
\label{eq:Lsugra}
\begin{split}
  {\cal L}_{4\rd} &= -\frac14 R +\frac12 g_{i \bar{j}} \? \partial_\mu z^i \partial^\mu \bar z^{\bar{j}} + \frac12\? (\partial_\mu \phi)^2 +  \frac{3\, e^{-4 \phi}}{16}\? H_{\mu\nu\rho} H^{\mu\nu\rho} \\
 &+ \frac12\, \left( \hat{\cal I}^{00}\?\hat{G}_{0, \mu\nu} \?
   \hat{G}^{\mu\nu}_0 + \hat{\cal
   I}_{UV}\? F_{\mu\nu}{}^U\? 
   F^{\mu\nu V} + 2\?\hat{\cal I}^0{}_U
   \?\hat{G}_{0, \mu\nu}\? F^{\mu\nu U} \right)    
 \\[.9ex]
  &{}
  + \frac14\, \varepsilon^{\mu\nu\rho\sigma} 
  \left( \hat{\cal R}^{00}\?\hat{G}_{0, \mu\nu} \?
   \hat{G}_{0, \rho\sigma} + \hat{\cal
   R}_{UV}\? F_{\mu\nu}{}^U\? F_{\rho\sigma}{}^V 
   + 2\?\hat{\cal R}^0{}_U    \?\hat{G}_{0, \mu\nu} \?
   F_{\rho\sigma}{}^U \right)    \\[.9ex]
& + \frac{1}8\? \varepsilon^{\mu\nu\rho\sigma}\? B_{\mu\nu} \left( s^2 F^1_{\rho \sigma} + s^1 F^2_{\rho \sigma} \right) + \frac12\? V (\phi, z, \bar{z}) \, ,
\end{split}
\end{eqnarray}
where $U, V = \{ 1, 2\}$, and the scalar potential still has the standard form,
\be
  \label{eq:potential}
  V=
  4\, L^{\Lambda} \bar L^{\Sigma}\? h_{uv}\,
  k^{u}{}\!_{\Lambda}\?k^{v}{}\!_{\Sigma} + \vec{P}_{\Lambda}\cdot\vec{P}_{\Sigma}\? \Big(g^{i\bar\jmath}
f_{i}{}^{\Lambda} \bar{f}_{\bar\jmath}{}^{\Sigma}
-3\?L^{\Lambda}\bar{L}^{\Sigma}\Big) \, ,
\ee
with $\Lambda, \Sigma = \{0, 1, 2 \}$, due to the purely electric gauging. 

This unusual duality frame can then be fully dualized back to the standard frame upon dualizing the tensor to a scalar and $G_0$ to $F^0$, which we perform next in order to keep the gauge kinetic matrices in a simple form for presentation purposes. We pause here to note that the form of \eqref{eq:Lsugra} is precisely the one we arrive at from the straightforward reduction ansatz we assumed in 6d. Notice that due to the truncation $\zeta = \tilde\zeta = 0$, we still have not fully fixed the Lagrangian as we have the freedom of adding arbitrary moment maps $P^{\U(1)}$ as explained above. They will be fixed in due course by matching the explicit scalar potential we find from the reduction with the general form in the above formula.

\subsection{Final dualization and scalar mapping}
Let us now perform a Legendre transformation to trade $\hat G_0$ for its electric counterpart $\hat F^0$. We shall extremize
\be
 \label{4d:eff:Leg}
 \cL_{4\rd}^{\text{eff, } \!e} \equiv \cL_{4\rd}^{\text{eff, } \!em}
 + \frac12 \varepsilon^{\mu \nu \rho \sigma} \hat F^0_{\mu \nu} \hat G_{0, \rho \sigma} \, ,
\ee
with respect to $\hat G_{0,\mu \nu}$,
\be
 \frac{\pd \cL_{4\rd}^{\text{eff, } \!em}}{\pd \hat G_{0,\mu \nu}} + \frac12 \varepsilon^{\mu \nu \rho \sigma} \hat F^0_{\rho \sigma} = 0 \, ,
\ee
and evaluate \eqref{4d:eff:Leg} at its critical point.
We find the standard $\cN=2$ gauged supergravity Lagrangian with two vector multiplets and a dualized hypermultiplet (also known as a tensor-scalar multiplet),
\bea
 \label{L4d:final}
  {\cal L}_{4\rd} &= -\frac14 R +\frac12 g_{i \bar{j}} \? \partial_\mu z^i \partial^\mu \bar z^{\bar{j}}
  + \frac12\? (\partial_\mu \phi)^2 + \frac12 \? \cI_{\Lambda \Sigma} \hat{F}_{\mu\nu}^{\Lambda} \hat{F}^{\mu\nu \Sigma}
  + \frac{1}{4} \varepsilon^{\mu\nu\rho\sigma}\? \cR_{\Lambda \Sigma} \hat{F}_{\mu\nu}^{\Lambda} \hat{F}_{\rho\sigma}^{\Sigma} \\
 &  +  \frac{3\, e^{-4 \phi}}{16}\? H_{\mu\nu\rho} H^{\mu\nu\rho} + \frac{1}8\? \varepsilon^{\mu\nu\rho\sigma}\? B_{\mu\nu} \left( s^2 F^1_{\rho \sigma} + s^1 F^2_{\rho \sigma} \right) + \frac12\? V (\phi, z, \bar{z}) \, ,
\eea
where the scalar potential $V$ is given by
\be
  V=
  4\? L^{\Lambda} \bar L^{\Sigma}\? h_{uv}\,
  k^{u}{}\!_{\Lambda}\?k^{v}{}\!_{\Sigma} + \vec{P}_{\Lambda}\cdot\vec{P}_{\Sigma}\? \Big(g^{i\bar\jmath}
f_{i}{}^{\Lambda} \bar{f}_{\bar\jmath}{}^{\Sigma}
-3\?L^{\Lambda}\bar{L}^{\Sigma}\Big) \, .  
\ee
Introducing the homogenous coordinates $X^\Lambda = (1 , z^i)$, $i=1,2$, our model is uniquely specified by the prepotential
\be
 \label{prepotential:reduced}
 F (X^\Lambda) = - \ii X^0 \sqrt{X^1 X^2} \, ,
\ee
and the quaternionic Killing vector and moment maps
\bea
 k^\bR & = \left\{ 0 ; -4m , s^2 , s^1 \right\} , \\
 P^1 & = P^2 = 0 \, , \qquad P^3 = \left\{ 0 ; 2 m e^{2 \phi} , 3 m - \frac{s^2}{2} e^{2 \phi} , 3 m - \frac{s^1}{2} e^{2 \phi} \right\} .
\eea
The vector multiplet scalar moduli space is a special K\"ahler manifold defined by the prepotential \eqref{prepotential:reduced}, and corresponds to the coset space $[\SU(1,1)/ \U(1)]^2$. A convenient parameterization of the complex scalars is
\be
 \label{z:X:chi}
 z^1 \equiv \frac{X^1}{X^0} = e^{2 \chi_1 + \chi_2} \, , \qquad z^2 \equiv \frac{X^2}{X^0} = e^{2 \chi_1 - \chi_2} \, .
\ee
We also define the symplectic covariant section
\be
 \cV = \{ L^\Lambda ; M_\Lambda \} \equiv e^{\cK / 2} \{ X^\Lambda ; F_\Lambda \} \, , \qquad F_\Lambda \equiv \frac{\pd F(X^\Lambda)}{\pd X^\Lambda} \, ,
\ee
from which we can construct
\be
 \label{f:h}
 f^\Lambda_i \equiv e^{\cK/2} (\pd_i + \pd_i \cK) X^\Lambda \, , \qquad  h_{\Lambda | i} \equiv e^{\cK/2} (\pd_i + \pd_i \cK) F_\Lambda \, .
\ee
Here, $\cK$ is the K\"ahler potential and reads
\be
 e^{- \cK} = \ii \left( \wb X^\Lambda F_\Lambda - X^\Lambda \wb F_{\Lambda} \right) = \left(e^{2 \chi_1 } + e^{2 \bar \chi_1} \right) ( 1 + \cosh ( \chi_2 - \bar \chi_2) ) \, .
\ee
The K\"ahler metric is then given by
\be
 g_{ i \bar j} = \pd_i \pd_{\bar j} \cK (\chi, \bar \chi) = \left(
\begin{array}{cc}
 \sech^2( \chi_1 - \bar \chi_1 ) & 0 \\
 0 & \frac{1}{1 + \cosh ( \chi_2 - \bar \chi_2 )} \\
\end{array}
\right) \, .
\ee
The period matrix reads
\be\label{eq:period}
 \cN_{\Lambda \Sigma} = \cR_{\Lambda \Sigma} + \ii \? \cI_{\Lambda \Sigma} = \wb F_{\Lambda \Sigma} + 2 \ii \? \frac{\im F_{\Lambda \Delta} \im F_{\Sigma \Theta} X^\Delta X^\Theta}{\im F_{\Delta \Theta} X^\Delta X^\Theta} \, .
\ee
We need the explicit values of $f^\Lambda_i$ to write down the scalar potential. Using \eqref{f:h} we obtain
\bea
 f^{0}_{1} & = - \frac{2 \sqrt{2} \? e^{\frac{1}{2} (4 \chi_1+\chi_2+\bar \chi_2)}}{\left(e^{2 \chi_1}+e^{2 \bar \chi_1}\right)^{3/2} \left(e^{\chi_2}+e^{\bar \chi_2}\right)} \, ,
 \qquad && f^{0}_{2} = \frac{\sqrt{2} \? e^{\frac{1}{2} (\chi_2+\bar \chi_2)} \left(e^{\bar \chi_2}-e^{\chi_2}\right)}{\sqrt{e^{2 \chi_1}+e^{2 \bar \chi_1}} \left(e^{\chi_2}+e^{\bar \chi_2}\right)^2} \, , \\
 f^{1}_{1} & = \frac{2 \sqrt{2} \? e^{\frac{1}{2} (4 \chi_1+3 \chi_2+4 \bar \chi_1+\bar \chi_2)}}{\left(e^{2 \chi_1}+e^{2 \bar \chi_1}\right)^{3/2} \left(e^{\chi_2}+e^{\bar \chi_2}\right)} \, ,
 \qquad && f^{1}_{2} = \frac{2 \sqrt{2} \? e^{\frac{3}{2} (\chi_2+\bar \chi_2)+2 \chi_1}}{\sqrt{e^{2 \chi_1}+e^{2 \bar \chi_1}} \left(e^{\chi_2}+e^{\bar \chi_2}\right)^2} \, , \\
 f^{2}_{1} & = \frac{2 \sqrt{2} \? e^{\frac{1}{2} (4 \chi_1-\chi_2+4 \bar \chi_1+\bar \chi_2)}}{\left(e^{2 \chi_1}+e^{2 \bar \chi_1}\right)^{3/2} \left(e^{\chi_2}+e^{\bar \chi_2}\right)} \, ,
 \qquad && f^{2}_{2} = - \frac{2 \sqrt{2} \? e^{\frac{1}{2} (4 \chi_1+\chi_2+\bar \chi_2)}}{\sqrt{e^{2 \chi_1}+e^{2 \bar \chi_1}} \left(e^{\chi_2}+e^{\bar \chi_2}\right)^2} \, .
\eea
We have checked that these quantities satisfy the following special geometry identities
\bea
 L^\Lambda \cI_{\Lambda \Sigma} \bar L^\Sigma & = - \frac12 \, , \qquad && L^\Lambda \cI_{\Lambda \Sigma} \bar f^\Sigma_i = 0 \, , \\
 f^\Lambda_i \cI_{\Lambda \Sigma} \bar L^\Sigma & = 0 \, , \qquad && f^\Lambda_i \cI_{\Lambda \Sigma} \bar f^\Sigma_{\bar j} = - \frac12 g_{i \bar j} \, , \\
 f^\Lambda_i g_{i \bar j} f^\Sigma_{\bar j} & = - \frac12 \cI^{\Lambda \Sigma} - \bar L^\Lambda L^\Sigma \, .
\eea
Finally, upon using the field redefinitions
\bea
 \chi_1 & = - \frac{1}{2} \log \Big( e^{2 ( C - \lambda) } - \frac{\ii}{2}b \Big) \, , \qquad \phi = - ( C + \lambda ) \, , \\
 \chi_2 & = \log \bigg( \frac{1}{\cosh \phi_0 \cosh \phi_3 - \sinh \phi_3} - \ii \frac{\sinh \phi_0}{\cosh \phi_0 - \tanh \phi_3} \bigg)
\, ,
\eea
one finds the Lagrangian \eqref{L4d:final} precisely matches the 4d effective one \eqref{eq:reducedLag}. We have explicitly checked this by first transforming the explicit gauge kinetic matrices appearing in \eqref{eq:reducedLag} evaluating $\hat G_0$ in terms of $\hat F^0$, and then matching these terms with the canonical expressions \eqref{eq:period} following from the parametrization \eqref{z:X:chi}.

\section{Consistent subtruncations in 4d}
\label{sec:trunc}

Having obtained the canonical $\cN = 2$ form of the reduced Lagrangian, we note that due to the gauging of the non-compact isometry we will always find one massive tensor multiplet (or massive vector multiplet in the standard duality frame) at the AdS$_4$ vacuum of the theory. This follows from the supersymmetry-preserving (\ie\,BPS) Stuckelberg (or Higgs) mechanism, leaving us with a single massless vector multiplet. As discussed in \cite{Hosseini:2017fjo,Hosseini:2020vgl} in the context of black hole near-horizon geometries, on the backgrounds that exhibit the BPS Higgs mechanism one can further describe the massless degrees of freedom using an effective prepotential. The effective prepotential is derived directly by imposing the BPS condition that $k^\mathbb{R}_\Lambda X^\Lambda$ and $k^\mathbb{R}_\Lambda A^\Lambda$ need to vanish, such that the remaining scalar manifold is described by
\be
\label{eq:effectiveF}
	F_{\text{eff}} = - \frac{\ii}{4 m} \left( s^2 X^1 + s^1 X^2 \right) \sqrt{X^1 X^2} \, .
\ee
One might however wonder if we can consistently truncate the massive modes and obtain a new supergravity theory, which means the massive modes would be integrated out at the level of the theory and not just decoupled on a given background. From the analysis of \cite{Hristov:2009uj,Hristov:2010eu} this can be achieved by imposing the maximally supersymmetric constraints on the massive modes {\it only}, while keeping free the remaining massless modes. This is in general {\it not} possible and is achieved only in special cases.

Following \cite{Hristov:2010eu} and \cite{Varela:2019vyd}, let us sketch the main constraints that ensure a consistent truncation can be achieved. We start from the theory in the standard duality frame with electric gaugings. The massive vector is integrated out using the hyperino variation condition  $\nabla_\mu \sigma = 0$ leading to the following integrability constraint,
\be
\label{eq:trunc1}
	k^\mathbb{R}_\Lambda A^\Lambda_\mu = 0  \qquad \Rightarrow \qquad  k^\mathbb{R}_\Lambda F^\Lambda_{\mu\nu} = 0\, ,
\ee 
 while the massive scalar in the vector multiplet is integrated out via
\be
\label{eq:trunc2}
	k^\mathbb{R}_\Lambda X^\Lambda = 0\, ,
\ee
again following from the hyperino variation. Let us denote the massive complex scalar that is fixed to a constant from the above equation as $\tau$. The last constraint is that the gaugino variation that corresponds to the scalar $\tau$ vanishes identically, which translates into
\be
\label{eq:maintruncationeq}
      g^{\tau \bar j} \bar f_{\bar j}^\Lambda \cI_{\Lambda \Sigma} F^\Sigma_{\mu\nu} = 0\, ,  \qquad   P^x_\Lambda f^\Lambda_\tau = 0\, , \quad  \forall x \in \{1, 2, 3 \}\, ,
\ee
with the second equation fixing uniquely the values of the remaining hypermultiplet scalars (apart from the free Goldstone boson, in this case $\sigma$, which is eaten by the massive vector). These two equations are non-trivial as they involve in general the full set of scalars and vectors and, depending on the model, might only be satisfied on a specific background. The truncation is therefore only fully consistent at the level of the theory when the last equations are identically solved for arbitrarily values of the remaining massless scalars and vectors. This means that apart from the constraints \eqref{eq:trunc1} and \eqref{eq:trunc2}, the remaining complex scalars and vectors should be kept arbitrary in \eqref{eq:maintruncationeq}. Instead, in case we want to truncate {\it all} the matter and keep only the gravity multiplet, we should impose \eqref{eq:maintruncationeq} for all the complex scalars. This will leave us with a single free gauge field (the graviphoton) and no free scalars.

If we perform these steps on the model we have identified in the previous section, we find two different consistent truncations that we describe in more detail below.

For completeness let us also sketch the same truncation as seen in the duality frame including the two-form field $B$ and its field strength $H$. Combining the condition  $\nabla_\mu \sigma = 0$ and the relation between the two frames found in \eqref{eq:Btosigma} we arrive at
\be
	H_{\mu\nu\rho} = 0\ ,
\ee
such that the equation of motion for $B$ following from \eqref{eq:Lsugra} becomes purely algebraic, 
\bea
\label{eq:eomforB}
	\hat{\cal I}^{00}\?\hat{G}_{0, \mu\nu} & + \hat{\cal I}^0{}_U\? F^U_{\mu\nu}    + \frac1{8 m}\? \varepsilon_{\mu\nu}{}^{\rho\sigma}\? \left( s^2 F^1_{\rho \sigma} +  s^1 F^2_{\rho \sigma} \right)
 \\[.9ex]
  & + \frac12\, \varepsilon_{\mu\nu}{}^{\rho\sigma} 
  \left( \hat{\cal R}^{00}\?  \hat{G}_{0, \rho\sigma} + \hat{\cal R}^0{}_U \?
   F_{\rho\sigma}{}^U \right)    = 0\, .
\eea
The remaining conditions spelled out above for the vectors and scalars remain the same, and we are then able to uniquely fix the two-form $B$ from the equation above. This procedure is equivalent to the one followed in 6d for fixing the two-form on BPS configurations \cite{Suh:2018tul,Hosseini:2018usu,Suh:2018szn}.%
\footnote{The form of \eqref{eq:eomforB} involving in general both $B$ and $*B$ in form notation does not allow us to write down a closed formula for $B$ itself. One can immediately compare for example that \cite[Eq.\,(5.3)]{Hosseini:2018usu} agrees with \eqref{eq:eomforB} in the case of vanishing axions considered there.}

\subsection[\texorpdfstring{$T^3$}{T**3} model]{$T^3$ model}

One of the subtruncations that are possible starting from the general 4d theory we obtained is the so-called gauged $T^3$ model.
It corresponds to a single vector multiplet with the scalar manifold $\SU(1,1)/\U(1)$ and constant FI gauging in the lack of hypermultiplets.
The same model comes out as a consistent subtruncation of the gauged $STU$ model that can be embedded in maximal 4d $\cN=8$ supergravity arising as a reduction of 11d supergravity on $S^7$ \cite{deWit:1986oxb,Cvetic:1999xp}.

In our case the $T^3$ model is obtained by setting%
\footnote{There is an obvious symmetry in our model under the exchange of indices $1$ and $2$, so we just pick one case without loss of generality.}
\be
\label{eq:additionaldataT3}
 \kappa = - 1 \, , \qquad s^1 = \frac{1}{3m} \, , \qquad s^2 = 0 \, ,
\ee
and the hyperscalar $\phi$ to its AdS$_4$ value
\be
  \mathring \phi = \frac12 \log (12 m^2) \, .
\ee
Furthermore, the supersymmetry preserving Higgs mechanism imposes
\be
 \label{Higgs:T3}
 k_\Lambda X^\Lambda = 0 \qquad \Rightarrow \qquad X^0 = \frac1{12 m^2} X^2 \qquad \xRightarrow{\eqref{z:X:chi}} \qquad \chi_2 = 2 ( \chi_1 - \mathring \phi ) \, ,
\ee
and thus we see that the massive complex scalar fixed to a constant (denoted as $\tau$ in the general discussion above) corresponds here to 
\be
	\tau = 2 \chi_1 - \chi_2\, , \qquad \Rightarrow \qquad \mathring \tau = \mathring \phi = \frac12 \log (12 m^2) \, ,
\ee
which allows us to verify successfully that \eqref{eq:maintruncationeq} is satisfied identically without additional constraints.

Using the constraint between the sections above, we can derive the effective prepotential of the reduced model parametrized by two of the original three sections,
\be
 F_{T^3} (X^I) = - \frac{\ii}{12 m^2} \sqrt{X^1 (X^2)^3} \, , \qquad I = 1, 2 \, .
\ee
We therefore need to effectively reduce all the symplectic vectors defining the model using the truncation rules above. This leads to identically vanishing Killing vectors and to a reduced form of the moment maps, which now have the meaning of constant FI terms after the hyperscalars have been truncated out,
\be
 \label{FI:G}
 P^3 \equiv G = \{ 0 , 0 ; 3 m , 3m \} \, .
\ee
Finally, after rescaling the field strengths as
\be
 F^0 = \frac{1}{12 m^2} F^2 = \wt F^2 \, , \qquad F^1 = \wt F^1 \, ,
\ee
we can bring \eqref{L4d:final} into the standard $\cN =2$ Lagrangian of the $T^3$ model,
\bea
 \label{LT3:final}
 \cL_{T^3} & = -\frac14 R \? + \frac{3}{2} \sech^2( \chi_1 - \bar \chi_1 ) \rd \chi_1 \rd \bar \chi_1
 - \frac{18 m^2 \left( 48 m^4 + e^{2 (\chi_1 + \bar \chi_1)} \right)}{e^{2 \chi_1} + e^{2 \bar \chi_1}} \\
 & + \frac12 \? \cI_{I J} \wt{F}_{\mu\nu}^{I} \wt{F}^{\mu\nu J} + \frac{1}{4} \varepsilon^{\mu\nu\rho\sigma}\? \cR_{I J} \wt{F}_{\mu\nu}^{I} \wt{F}_{\rho\sigma}^{J} \, .
\eea
with the FI gauging \eqref{FI:G} and the following special geometry data. The K\"ahler potential, up to K\"ahler transformations, reads
\be
 e^{- \cK} = 4 e^{\chi_1 + \bar \chi_1} \cosh^3 ( \chi_1 - \bar \chi_1) \, ,
\ee
where we used
\be
	\frac{X^2}{X^1} = (12 m^2)^2\? e^{- 4 \chi_1}\, ,
\ee
to parametrize the same scalar $\chi_1$ that we encountered in the bigger model.
The K\"ahler metric is then given by
\be
g_{1 \bar 1} = 3 \sech^2 ( \chi_1 -\bar \chi_1 ) \, .
\ee
The scalar potential has the standard form
\be
 V_{T^3} (\chi_1 , \bar \chi_1) = g_I g_J \left( g^{1 \bar 1} f^I_1 \bar f^J_{\bar 1} - 3 \bar L^I L^J \right) ,
\ee
with
\bea
 f^1_1 & = -\frac{6 m^2 e^{- \frac{1}{2} ( \chi_1 + \bar \chi_1)} (3 \tanh (\chi_1-\bar \chi_1)+1)}{\cosh ^{\frac{3}{2}}(\chi_1-\bar \chi_1)} \, , \\
 f^2_1 & = \frac{e^{\frac{1}{2} (5 \chi_1 + \bar \chi_1)}}{8 m^2 \cosh ^{\frac{5}{2}}(\chi_1-\bar \chi_1)} \, .
\eea

We  have thus found a string theory embedding for the 4d gauged $T^3$ model in type IIA/IIB via the passage through 6d F(4) supergravity on $\Sigma_\fg$.
As mentioned in the beginning of this subsection, the same $T^3$ model can be embedded in 11d supergravity on $S^7$
and therefore one can find numerous results in the literature concerning supersymmetric solutions \cite{Cacciatori:2009iz,Freedman:2013oja,Katmadas:2014faa,Halmagyi:2014qza,Hristov:2018spe,Hristov:2019mqp,Hosseini:2019and,Bobev:2020pjk}
and even thermal black holes in this theory \cite{Duff:1999gh,Klemm:2012yg,Toldo:2012ec}.
We are then free to take these results and interpret them now also in the type IIA/IIB setting.
We give a few examples of this when discussing rotating black holes in section \ref{subsec:rotatingBH}.

\subsection{Minimal gauged supergravity}
\label{sect:minimal}

Notice that in the above truncation we specified from the beginning the value of the a priori free parameters $s^{1,2}$.
If we want to find a subtruncation while still allowing for an arbitrary $\kappa$ and $s^{1,2}$ it turns out we cannot keep any of the additional matter multiplets.
The reason is that \eqref{eq:maintruncationeq} can only be solved by fixing all scalars to their AdS$_4$ values if we do not commit to \eqref{eq:additionaldataT3},
\bea
 \mathring \chi_1 & = \frac{1}{4} \log \left(\frac{96 m^3}{m \left(9 (s^ 1 -s^2 )^2 + 12 s^1 s^2 \right) - \kappa \sqrt{ 9 (s^1 - s^2 )^2 + 4 s^1 s^2}} \right) , \\
 \mathring \chi_2 & = \frac{1}{2} \log \left(\frac{2 s^1}{\sqrt{9 (s^1-s^2 )^2+4 s^1 s^2}+3 ( s^1 - s^2 )}\right) , \\
 \mathring \phi & = \frac{1}{2} \log \left( \frac{24 m^2}{- \kappa + m \sqrt{9 (s^1-s^2 )^2 + 4 s^1 s^2}}\right) ,
\eea
where we have assumed that $s^1$ is never vanishing.
It is well known that any matter-coupled supergravity admits a consistent truncation to minimal gauged supergravity, and in our case this is achieved by further identifying
\be
	F^0 = e^{-\mathring \chi_1}\, \wt{F}\, ,  \qquad F^1 = e^{\mathring \chi_1 + \mathring \chi_2}\, \wt{F}\, , \qquad F^2 = e^{\mathring \chi_1 - \mathring \chi_2}\, \wt{F}\,  .
\ee
The relation between $F^0, F^1$ and $F^2$ above is fixed uniquely by supersymmetry since the gaugino variations require the combinations $f_i^\Lambda \cI_{\Lambda \Sigma} F^\Sigma$ to vanish.
The overall normalization concerning the rescaled field strength $\wt{F}$ was chosen such that one can derive the truncated theory from the prepotential
\be
 \label{prepotential:minimal}
	F_\text{min} = -\ii\? \wt{X}^2\, .
\ee
The corresponding Lagrangian describes only the gravity multiplet of $\cN = 2$ supergravity and its bosonic part is equivalent to Einstein-Maxwell theory with a cosmological constant,
\be
	 \label{eq:minsugra}
 \cL_\text{min} = -\frac14 R \? -   \wt{F}_{\mu\nu} \wt{F}^{\mu\nu} + \frac12 \? V_{\text{min}} \, .
\ee
The scalar potential is given by
\be
 \label{V:minimal}
 V_{\text{min}} = - \frac{3}{L^2_{\text{AdS}_4}} = - 2 \? ( 6 m)^4 \? \frac{\sqrt{2 \kappa  \left( \kappa -\sqrt{\kappa ^2+8 z^2} \right)+4 z^2}}{\left(\sqrt{\kappa ^2+8 z^2}-3 \kappa \right)^2} \, ,
\ee
such that the gauge coupling constant entering the standard gauged supergravity formulation is given by $g = 2 L_{\text{AdS}_4}^{-1}$.
Here, the variable $z$ parametrizes the fluxes
\be
 s^1 = - \frac{\kappa}{6 m} \left( 1 + \frac{z}{\kappa } \right) , \qquad s^2 = - \frac{\kappa}{6 m} \left( 1 - \frac{z}{\kappa} \right) .
\ee

Let us emphasize that in the above truncation we only kept the graviphoton, which is considered dual to the $\U(1)$ R-current, and the solutions of this theory can be embedded into string or M-theory in various ways, see \cite{Gauntlett:2007ma,Larios:2019lxq}.
However, in contrast with the \emph{universal} solutions presented in \cite[Sect.\,4.2]{Bobev:2017uzs}, our more general backgrounds depend on the twist parameter $z$ for the $\U(1) \subset \SU(2)_M$ flavor symmetry.
Setting $\kappa = -1$ and $z=0$ we find, upon uplift to minimal 6d F(4) supergravity, the AdS$_4 \times \Sigma_{\fg>1}$ vacuum of \cite{Naka:2002jz}.
The scalar potential in this case reads
\be
 V_{\text{min}} \Big|_{\kappa = -1 , \, z=0} = - 4 (3 m)^4\, .
\ee

\section[Universal AdS\texorpdfstring{$_4$}{(4)} solutions]{Universal AdS$_4$ solutions}
\label{sec:universal}

We are now in a position to discuss some interesting solutions of the 4d theory we have found. We can first focus on the simplest case of the subtruncation to minimal supergravity. In this case a multitude of asymptotically AdS$_4$ solutions are already known, see \cite{Caldarelli:1998hg,Chamblin:1998pz,AlonsoAlberca:2000cs} and many consequent references. More recently, a simple organizing principle was put forward in \cite{BenettiGenolini:2019jdz} for all supersymmetric asymptotically AdS$_4$ solutions in terms of fixed points (NUTs) or fixed two-manifolds (Bolts) of the isometry arising from the Killing spinor on the given background. This allows us to write down a general expression for the holographically renormalized on-shell action of all supersymmetric solutions. We refer the reader for more details to \cite{BenettiGenolini:2019jdz}, while here we can make a simple example to illustrate the idea. For a solution with a number of fixed points, the resulting on-shell action reads
\be
\label{eq:NUTs}
	I = \frac{\pi L^2_{\text{AdS}_4}}{2 G_{\text{N}}^{4\rd}} \sum_{\text{NUTs}} \frac{(\epsilon+1)^2}{4\? \epsilon}\, ,
\ee
where $L^2_{\text{AdS}_4}$ is given in \eqref{V:minimal}, and at each isolated fixed point one encounters an omega deformation parameter $\pm \epsilon$.
In general, such solutions correspond to Euclidean AdS$_4$ geometries with squashed $S^3$ boundary, or to rotating black holes in AdS$_4$.
Holographically they relate to the squashed $S^3$ free energy or superconformal indices of the dual field theories.
Below we discuss in some more detail the free energy of the dual SCFTs on a sphere, while we devote the next sections to the black hole cases,
which we present not only in the minimal subtruncation but in the more general matter-coupled 4d theory.
We do not discuss at further length the Bolt solutions, but we note that all supergravity results in \cite{Toldo:2017qsh} can be applied to our case
and there is an analogous formula that generalizes \eqref{eq:NUTs} to include the Bolts.
It would be interesting to further generalize all such minimal supergravity solutions to the full matter-coupled model we found above.

Let us for the sake of clarity stress again that the results in 4d minimal supergravity should not be considered as equivalent to those in minimal 6d F(4) supergravity
since the 4d theory incorporates the effect of the additional $\U(1)$ gauge field for the $\SU(2)_M$ flavor symmetry.
In practice, this for example means that our result for the $S^3$ free energy of the class $\cF$ theories is more general than the answer coming from the \emph{universal} flow from 6d in \cite{Bobev:2017uzs},
owing to the additional parameter $z$.

\subsection[The \texorpdfstring{$S^3$}{S**3} free energy]{The $S^3$ free energy}

The AdS$_4$ vacuum described in section \ref{sect:minimal} can be uplifted to 10d massive type IIA and IIB supergravities.
The complete spacetime can be thought of as interpolating between AdS$_6 \times M_4$ and AdS$_4 \times \Sigma_{\fg} \times M_4$ vacua,
leading to a natural holographic interpretation as a renormalization group (RG) flow across dimensions:
we have a flow from the 5d $\cN = 1$ SCFT compactified on $\Sigma_\fg$ to a 3d SCFT.
In general, we have a family of such 3d SCFTs labeled by a set of magnetic fluxes, parameterizing the twist, that we call \emph{class $\cF$ theories}.
The free energy of class $\cF$ theories on $S^3$ can be computed directly using \eqref{V:minimal},
or alternatively from the general squashed sphere free energy with a single fixed point in \eqref{eq:NUTs} in the $\epsilon \to 1$ limit (no squashing).
Setting $m = \frac12$ such that in the parent theory we find $( L_{\text{AdS}_6} = 1)$, we find
\be
 \label{FS3:holo:classF}
 F_{ S^3} = \frac{\pi L^2_{\text{AdS}_4}}{2 G_{\text{N}}^{4\rd}}
 = - \frac19 | 1 - \fg | F_{S^5} \? \frac{\left(\sqrt{\kappa ^2+8 z^2}-3 \kappa \right)^2}{\sqrt{2 \kappa  \left( \kappa -\sqrt{\kappa ^2+8 z^2} \right)+4 z^2}} \, ,
\ee
where we used the standard AdS$_6$/CFT$_5$ dictionary,
\be
 \label{holo:dict}
 \frac{1}{G_\text{N}^{6\rd}} = - \frac{3}{\pi^2} F_{S^5} \qquad \xRightarrow{\eqref{volSigma}} \qquad
 \frac{1}{G_{\text{N}}^{4\rd}} = \frac{\vol (\Sigma_\fg)}{G_\text{N}^{6\rd}} = - \frac{12 | 1 - \fg |}{\pi} \? F_{S^5} \, .
\ee
Here, $F_{S^5}$ is the free energy of the \emph{parent} $\cN = 1$ SCFTs on $S^5$.

For Seiberg $\USp(2N)$ gauge theories with $N_f$ fundamental flavors and an antisymmetric matter field, arising on the worldvolume of D4-branes near D8-branes and O8-planes \cite{Seiberg:1996bd},
the $S^5$ free energy was computed in \cite{Jafferis:2012iv} and it is give by
\be
 \label{F:S5}
 F_{S^5}^{\text{Seiberg}} = - \frac{9 \sqrt{2} \? \pi}{5} \frac{N^{5/2}}{\sqrt{8 - N_\text{f}}} \, .
\ee
The $S^3$ free energy of this category of class $\cF$ theories can then be explicitly written as
\be
 \label{FS3:holo}
 F_{ S^3}^{\text{Seiberg}} = \frac{\sqrt{2} \? \pi | 1 - \fg | }{5} \frac{N^{5/2}}{\sqrt{8 - N_\text{f}}} \? \frac{\left(\sqrt{\kappa ^2+8 z^2}-3 \kappa \right)^2}{\sqrt{2 \kappa  \left( \kappa -\sqrt{\kappa ^2+8 z^2} \right)+4 z^2}} \, ,
\ee
in complete agreement with the free energy derived directly in 10d massive type IIA supergravity \cite[Eq.\,(1.1)]{Bah:2018lyv} and the field theory computations of \cite{Crichigno:2018adf,Hosseini:2018uzp}.

A larger class of 5d SCFTs can be engineered using $(p,q)$ 5-branes in type IIB supergravity, among them $T_N$ and $\#_{M,N}$ theories.
The $T_N$ theories are realized on an intersection of $N$ D5-branes, $N$ NS5-branes and $N$ $(1,1)$ 5-branes \cite{Benini:2009gi}.
Interestingly, upon a circle reduction they become the 4d $T_N$ theories of \cite{Gaiotto:2009we}.
The $\#_{M,N}$ theories are realized on the intersection of $N$ D5-branes with $M$ NS5-branes \cite{Aharony:1997bh}.
The free energy of this class of field theories on $S^5$ was computed in \cite{Fluder:2018chf,Uhlemann:2019ypp} and they read
\be
 F_{S^5}^{T_N} = - \frac{27 \zeta ( 3)}{8 \pi^2} N^4 \, , \qquad F_{S^5}^{\#_{M,N}} = - \frac{189 \zeta ( 3)}{16 \pi^2} N^2 M^2 \, .
\ee
The $S^3$ free energy of the corresponding class $\cF$ theories can thus be written as
\bea
 \label{F:S3:TN}
 F_{ S^3}^{T_N} & = \frac{3 \zeta ( 3) | 1 - \fg |}{8 \pi^2} N^4 \? \frac{\left(\sqrt{\kappa ^2+8 z^2}-3 \kappa \right)^2}{\sqrt{2 \kappa  \left( \kappa -\sqrt{\kappa ^2+8 z^2} \right)+4 z^2}} \, , \\
 F_{ S^3}^{\#_{M,N}} & = \frac{21  \zeta ( 3) | 1 - \fg |}{16 \pi^2} N^2 M^2 \? \frac{\left(\sqrt{\kappa ^2+8 z^2}-3 \kappa \right)^2}{\sqrt{2 \kappa  \left( \kappa -\sqrt{\kappa ^2+8 z^2} \right)+4 z^2}} \, .
\eea
It would be interesting to derive \eqref{F:S3:TN} by taking the large $N$ limit of the $S^3 \times \Sigma_\fg$ partition function of $T_N$ and $\#_{M,N}$ theories.
This can be done along the lines of \cite{Uhlemann:2019ypp}.

It is clear that we can analogously derive the same type of holographic predictions for squashed spheres \cite{Martelli:2011fu,Martelli:2011fw} and other supersymmetric solutions using the general formula \eqref{eq:NUTs} and its generalization with Bolts.
The same considerations can also lead us to the supersymmetric R\`enyi entropy of the class $\cF$ theories \cite{Nishioka:2013haa,Nishioka:2014mwa,Huang:2014gca}.

\section{Black hole solutions}
\label{sec:BlackHoles}

In this section we write down supersymmetric black hole solutions of the 4d F(4) gauged supergravity. We find some genuinely new static dyonic black holes to the general 4d model with hypermultiplet gaugings derived above, generalizing the purely magnetic ones found directly in 6d in \cite{Hosseini:2018usu,Suh:2018szn}. We also write down rotating black holes with and without a twist in the $T^3$ subtruncation with FI gauging. Even though the solutions of the $T^3$ model are not new, our reduction provides a novel embedding in string theory. This in turn leads to new holographic predictions
as we discuss in the next section. 
 
All the BPS solutions we discuss here are derived by solving the first order integrability conditions following from the existence of Killing spinors generating a timelike isometry, derived in detail in \cite{Cacciatori:2008ek,Meessen:2012sr,Chimento:2015rra} in a duality frame with purely electric gauging. These conditions were further simplified and solved explicitly in a series of related papers for the various classes of black holes we consider here. 

Static magnetic black holes in AdS$_4$ were first found in \cite{Cacciatori:2009iz} for models with FI gauging, and their dyonic generalizations were understood in \cite{Katmadas:2014faa,Halmagyi:2014qza}. The problem of generalizing those solutions to cases with hypermultiplet gaugings was discussed first in \cite{Halmagyi:2013sla,Chimento:2015rra} and gradually more examples were added in \cite{Amariti:2016mnz,Guarino:2017pkw,Hosseini:2017fjo,Benini:2017oxt,Bobev:2018uxk,Benini:2020gjh} based on the effective decoupling of the massive vector near the horizon. 

The addition of rotation to the twisted black holes described above, as well as in the case of KN-AdS without a twist, is still an open problem for the models with general hypermultiplet gauging.
A first example of this kind, to the best of our knowledge, was given recently in \cite{Hosseini:2020vgl}, based on a generalization of the rotating solutions with FI gaugings in \cite{Hristov:2018spe},
\cite{Hristov:2019mqp} and \cite{Hosseini:2019lkt} for rotating twisted AdS$_4$ black holes, KN-AdS$_4$ black holes, and twisted rotating AdS$_5$ black strings, respectively. 
In order to proceed with the explicit solutions in the different cases, we should first introduce the so-called $I_4$-formalism, which is defined for all symmetric special K\"ahler models. In such cases one can define the invariant contraction of up to four different symplectic vectors, such as the symplectic section $\{X^\Lambda ; F_\Lambda \}$ and the vector of electromagnetic charges
$ \Gamma = \{ p^\Lambda ; q_\Lambda \}$. The invariant contraction is called the {\it quartic} invariant, or $I_4$, see the appendices of \cite{Bossard:2013oga,Halmagyi:2014qza,Hristov:2018spe,Hosseini:2019iad} for further definitions and useful identities involving symplectic inner products $\Iprod{\cdot}{\cdot}$ and the $I_4$. Here we just note that for our model defined by the prepotential
\be
	F =- \ii X^0 \sqrt{X^1 X^2} \, ,
\ee
the quartic invariant is given by
\be
 I_4 ( \Gamma ) = 2 p^0 p^1 q_0 q_1+  2 p^0 p^2 q_0 q_2 - ( p^1 q_1 - p^2 q_2 )^2 + ( p^0 )^2 p^1 p^2 + 4 q_0^2 q_1 q_2 \, ,
\ee
which is enough to define uniquely all the rest of the quantities we discuss in the specific examples below.

\subsection[Static magnetic black holes of \texorpdfstring{$\cite{Hosseini:2018usu}$}{[1]}]{Static magnetic black holes of \cite{Hosseini:2018usu}}
\label{subsec:knownbh}
We first look at the case of static magnetic black holes with a twist, and directly zoom in on the near-horizon region of the product form AdS$_2 \times \Sigma_\fg$. Upon uplift to six dimension via the formulae in the previous sections, one recovers the near-horizons of the form AdS$_2 \times \Sigma_{\fg_1} \times \Sigma_{\fg_2}$ in \cite[Sect.\,5]{Hosseini:2018usu}. The full flow from the near-horizon to the asymptotic AdS$_6$ was constructed numerically in \cite{Suh:2018szn}. From four-dimensional perspective, these can be found most directly in the duality frame with a hypermultiplet using the equations of \cite{Chimento:2015rra}. One can see a more detailed account of how these equations are solved step by step in \cite{Hosseini:2017fjo}, here we just present the main steps.

We only include magnetic charges here, and the truncation of the massive vector at the near-horizon translates into the condition
\be
\label{K:constr}
	 \Iprod{k^\bR}{\Gamma}  = 0 \, ,
\ee
which means that
\be
 \Gamma = \left\{ \frac{1}{4 m} ( p^2 s^1 + p^1 s^2 ),  p^1 , p^2 ; 0 \right\} .
\ee
The magnetic charges further satisfy the twisting condition
\be
 \label{T:const:p}
 \Iprod{P^3}{\Gamma}  = - \kappa \quad \Rightarrow \quad p^1 + p^2 = - \frac{\kappa}{3 m} \, ,
\ee
with $\kappa$ being the curvature of the Riemann surface. The scalars at the near-horizon are fixed to constants and in the symplectic formalism are parametrized by
\be
 \cH_0 = \left\{ \frac{1}{4 m} ( a_2 s^1 + a_1 s^2 ), a_1, a_2 ; 0 \right\} ,
\ee
where we have already imposed that $\Iprod{k^\bR}{\cH_0} = 0$. The value of the sections can be then recovered by
\be
\label{eq:staticHtosection}
	\{ X^\Lambda; F_\Lambda \} = - \frac{1}{2 \Iprod{P^3}{\cH_0} \sqrt{I_4 (\cH_0)}}\? I_4' (\cH_0) + \frac{\ii}{\Iprod{P^3}{\cH_0}}\? \cH_0\ .
\ee
The value of the hypermultiplet scalar $\phi$ is fixed from the condition $I_4 (k^\bR, P^3, \cH_0, \cH_0) = 0$, which arises from compatibility of the attractor equations with \eqref{K:constr},
\be
 e^{2 \phi} = \frac{12 m ( a_1^2 s^2 + a_2^2 s^1 )}{3 a_2^2 ( s^1 )^2 + 2 a_1 a_2 s^1 s^2 + 3 a_1^2 ( s^2 )^2} \, .
\ee
The attractor equations themselves relate the charges $\Gamma$ and the vector $\cH_0$, thus fixing the scalars in terms of charges. We find the relations
\bea
	p^1 &= \frac{3 m a_1 (a_1 s^2+a_2 s^1) (a_1 s^2 (a_1 + 3 a_2) + 3 a_2 s^1 (a_2-a_1))}{3 a_2^2 ( s^1 )^2 + 2 a_1 a_2 s^1 s^2 + 3 a_1^2 ( s^2 )^2}\, , \\  p^2 &= \frac{3 m a_2 (a_1 s^2+a_2 s^1) (3 a_1 s^2 (a_1-a_2) + a_2 s^1 (3 a_1 + a_2))}{3 a_2^2 ( s^1 )^2 + 2 a_1 a_2 s^1 s^2 + 3 a_1^2 ( s^2 )^2}\, ,
\eea
and for brevity we choose not to explicitly present the form of the parameters $a_{1,2}$ in terms of $p^{1,2}$ and $s^{1,2}$. Note that due to the constraint among the charges, only one of the two parameters $a_{1,2}$ is free.
Finally, the black hole entropy given by the Bekenstein-Hawking formula, reads 
\be
 S_{\text{BH}} = \frac{\text{Area}}{4 G^{4\rd}_{\text{N}}} = \frac{\vol (\Sigma_\fg)}{16 m G_{\text{N}}^{4\rd}} ( a_1 s^2 + a_2 s^1 ) \sqrt{a_1 a_2} \, ,
\ee
and implicitly is a function of the magnetic fluxes through the Riemann surface $p^{1,2}$ corresponding to a one-parameter family of solutions (if we assume the 4d point of view where the parameters $s^{1,2}$ are part of the definition of the theory).

\subsection{General static dyonic black holes}
\label{subsec:staticBH}

Here, we add electric charges to the static magnetic solutions presented above. We follow the same steps as above but additionally need to impose the vanishing NUT condition
\be
	\Iprod{P^3}{I_4'(\cH_0)} = 0\, .
\ee
This constraint is automatically satisfied in the absence of electric charge but here imposes an additional non-linear constraint among the charges.
Taking into account that the massive vector does not give rise to a conserved charge, one can think of the solution presented below as the addition of a single free electric charge on top of the purely magnetic solution.  The charge vector is parametrized by
\be
 \Gamma = \left\{ \frac{1}{4 m} ( p^2 s^1 + p^1 s^2 ),  p^1 , p^2 ; q_0, q_1, q_2 \right\} ,
\ee
satisfying the same twisting condition \eqref{T:const:p}. The scalars are parametrized now by the parameters $a_{1,2}$ and $b$, such that
\be
 	\cH_0 = \left\{ \frac{a_2 s^1 + a_1 s^2 }{4 m}, a_1, a_2 ; b, \frac{b}{8 m}\? \left( s^2 + \frac{c_2}{d_1}\? s^1 \right) , \frac{b}{8 m}\? \left( s^1 + \frac{c_1}{d_2}\? s^2 \right)\right\} ,
\ee
where for convenience here and in later expressions we define the short-hand notation
\bea
	c_1 &= a_1^2 + 3 a_1 a_2 + 2 b^2\, , \qquad \qquad c_2 = a_2^2 + 3 a_1 a_2 + 2 b^2\, ,\\ 
	d_1 &=  a_1^2 - a_1 a_2 - 2 b^2\, , \qquad \qquad d_2 = a_2^2 - a_1 a_2 - 2 b^2\, , \\ 
	f_1 &= 3 a_1^2 + a_1 a_2 - 2 b^2\, , \qquad \qquad f_2 = 3 a_2^2 + a_1 a_2 - 2 b^2\, ,\\ 
	\Pi &= 3 d_2^2 (s^1)^2 + 3 d_1^2 (s^2)^2 + 2 s^1 s^2 (  a_1 a_2 (a_1-a_2)^2 + 4b^2 (a_1+a_2)^2 - 4 b^4)\, .
\eea
The value of the hypermultiplet scalar $\phi$ is then fixed to
\be
 e^{2 \phi} = \frac{12 m ( d_1^2 s^2 + d_2^2 s^1 )}{\Pi} \, .
\ee
The attractor mechanism fixes the scalars in terms of the charges via
\bea
	p^1 &= \frac{3 m\? d_1}{d_2\? \Pi}\? (s^2 d_1 - s^1 d_2)\? (3 s^1 d_2^2 - s^2 d_1 f_2)\, , \\
	p^2 &= \frac{3 m\? d_2}{d_1\? \Pi}\? (s^1 d_2 - s^2 d_1)\? (3 s^2 d_1^2 - s^1 d_2 f_1)\, ,
\eea
and
\bea
	q_0 &= \frac{3 m\? b\? (a_1+a_2)}{d_1 d_2\? \Pi}\? (s^2 d_1 - s^1 d_2)\? \left(s^1 d_2^2 (f_1 -2 a_1 a_2 -2 b^2) - s^2 d_1^2 (f_2 -2 a_1 a_2 -2 b^2) \right)\, \\
	q_1 &= \frac{3 b\? (a_1+a_2)}{8 d_1 d_2\? \Pi}\? (s^2 d_1 - s^1 d_2)\? \times \Big[  8 s^1 s^2 (d_1+d_2) (a_2^2-b^2) (a_1 a_2 + b^2)\\
& \qquad \quad \qquad + 3 (s^1)^2 d_2^2 (d_1-2 a_1 a_2 - 2 b^2) - (s^2)^2 d_1^2 (f_2 - 2 a_1 a_2 - 2 b^2) \Big]\, , \\
	q_2 &= \frac{3 b\? (a_1+a_2)}{8 d_1 d_2\? \Pi}\? (s^2 d_1 - s^1 d_2)\? \times \Big[ 8 s^1 s^2 (d_1+d_2) (a_1^2-b^2) (a_1 a_2 + b^2)  \\
& \qquad \quad \qquad + 3 (s^2)^2 d_1^2 (d_2-2 a_1 a_2 - 2 b^2) - (s^1)^2 d_2^2 (f_1 - 2 a_1 a_2 - 2 b^2) \Big]\ \, .
\eea
We will not attempt to write the parameters $a_{1,2}$ and $b$ in terms of the charges, but we note that due to the twisting condition \eqref{T:const:p} only two of these parameters are independent. The black hole entropy is then given by
\be
 S_{\text{BH}} = \frac{\vol (\Sigma_\fg)}{16 m\?  G_{\text{N}}^{4\rd}}\?  \frac{(s^2 d_1 - s^1 d_2)}{d_1 d_2}\? \sqrt{\left( (a_1 -a_2)^2-4 b^2 \right) \left(a_1 a_2+b^2\right)^3}\, .
\ee
We expect that there exist numerical solutions interpolating between the full parameter space of near-horizon solutions presented here and the asymptotic AdS$_4$ region, as well as another flow connecting the same near-horizon to AdS$_6$ when embedded in six-dimensional supergravity. We leave the explicit construction of such solutions as a future research direction.

\subsection[Rotating black holes in the \texorpdfstring{$T^3$}{T**3} limit]{Rotating black holes in the $T^3$ limit}
\label{subsec:rotatingBH}

We can also discuss cases with rotation, for which we need to specify our model to the $T^3$ truncation. The rotating solutions for models with general hypermultiplet gaugings are yet to be constructed and are beyond the scope of the present work. In order to keep the symplectic vectors to be the same size as above for a clear comparison, we will not use the resulting $T^3$ prepotential, but only the specification
\be
 \kappa = - 1 \, , \qquad s^1 = \frac{1}{3m} \, , \qquad s^2 = 0 \, .
\ee
At the level of BPS equations this is really the same as looking at the $T^3$ model. We will again only look at the near-horizon geometries, but we note that due to the constant hypermultiplet scalars this subtruncation allows for a completely analytic form of the complete black hole spacetime as presented in the original references we give.

\subsubsection{Twisted black holes in AdS}

We first look at the case of the twisted black holes discussed above, generalizing the magnetic solutions to include rotation. We insist on the horizon to remain compact, which means we can only look at the spherical case. The near-horizon geometry then becomes a fibration of a squashed $S^2$ over AdS$_2$. The magnetic rotating solution of the $T^3$ model was constructed explicitly in \cite[Sect.\,4.3.1]{Hristov:2018spe} and further discussed carefully in \cite[Sect.\,2.2]{Hosseini:2019iad}.
Here we just list the main formluae characterizing the solution, setting $m = \frac12$ for simplicity. The vector of electromagnetic charges is given by
\be
 \Gamma = \left\{ -\frac{1}{3} \left(  \frac23 + p^1 \right) \! ,  p^1 , -\frac23 - p^1 ; 0 \right\} ,
\ee
where we already implemented the twisting condition \eqref{T:const:p} and the specialization to $s^2 = 0$. After imposing the a number of constraints in analogy to the cases above, we find the scalars parametrized simply by
\be
 \cH_0 = \left\{ \frac{a_2}3, a_1, a_2 ; 0 \right\}  .
\ee
Due to the rotation, the scalars are no longer constants at the horizon and depend on one of the angular directions on the sphere via the generalization of \eqref{eq:staticHtosection},
\be
\label{eq:Htosection}
	\{ X^\Lambda; F_\Lambda \} = - \frac{I_4' (\cH_0 + j P^3 \cos \theta)}{2 \Iprod{P^3}{\cH_0} \sqrt{I_4 (\cH_0 + j P^3 \cos \theta)}}  + \frac{\ii}{\Iprod{P^3}{\cH_0}}\? (\cH_0 + j P^3 \cos \theta) \, ,
\ee
where the parameter $j$ is related to the conserved angular momentum $\cJ$ via the attractor equations.
They now fix the parameters $a_{1,2}$ and $j$ in terms of the magnetic charge $p^1$ and the conserved angular momentum $\cJ$,
\bea
	p^1 + \frac{3^5}{(6 p^1+1)^3}\, \cJ^2 &= \frac32\? a_1 (a_2 -a_1)\, ,\\
	-p^1 + \frac{3^6}{(6 p^1+1)^3}\, \cJ^2 &= \frac23 + \frac12\? a_2 (3 a_1 + a_2)\, ,\\
	\frac{3^3}{(6 p^1+1)^3}\, \cJ^2 &= - j^2\, .
\eea
The Bekenstein-Hawking entropy is given by
\be
 S_{\text{BH}} = \frac{\pi}{9 \sqrt{2} \? G_{\text{N}}^{4\rd}} \sqrt{ 1 - 6 p^1 (3 p^1+1) - \sign(6 p^1+1) \sqrt{(2 p^1+1) (6 p^1+1)^3 - 4 \times 3^5 \cJ^2} } \, ,
\ee
with regular solutions existing for $p^1< 0$, and one can see that the absolute value of the angular momentum is bounded from above. 

\subsubsection{Kerr-Newman black holes in AdS}

Let us now look at the Kerr-Newman solution, which in the $T^3$ model contains only electric charges and rotation. The near-horizon geometry can again only allow for a spherical topology and becomes a fibration of a squashed $S^2$ over AdS$_2$. This solution was explicitly written in \cite{Hristov:2019mqp} and therefore we keep the discussion here short. We again set $m = \frac12$ for simplicity, and have the electromagnetic charge vector given by
\be
	\Gamma = \{ 0; 6 q_2, q_1, q_2 \}\, .
\ee
The scalars in this case are parametrized by the vector
\be
	\cC = \left\{0; b_0, b_1, \frac16\? b_0 \right\} ,
\ee 
where the precise relation between the sections and the above vector is no longer given by \eqref{eq:Htosection}, but by a more complicated expression that the interested reader can find in \cite{Hristov:2019mqp}.
In analogy to the case above, the scalars at the horizon are functions of the angular coordinate $\theta$ on the sphere.
The attractor equations relate the conserved charges $q_{1,2}$ and $\cJ$ to the parameters $b_{0,1}$ such that we find a two-parameter family of solutions.
Explicitly we obtain%
\footnote{We have an overall sign difference with the definition of angular momentum in \cite{Hristov:2019mqp}. We have chosen this convention to present in a unifying way the entropy functions in the next section.}
\bea
q_1 &= \frac{1}{4 \Xi}\? \left( 4 b_1 -3 b_0^3 + 18 b_0^2 b_1 \right)  ,\quad
q_2 = \frac{1}{12 \Xi}\? \left(2 b_0 + 3 b_0^3 + 6 b_0^2 b_1 \right) ,\\
\cJ &= -\frac{1}{2 \Xi^2}\? b_0^2\? \left(b_0 + 6 b_1+6 b_0 b_1 (4 b_1+3 b_0 (2 + b_0^2 + 6 b_0 b_1)) \right) ,
\eea
where we defined $\Xi \equiv 1 - 18\? b_0^3\? b_1$. In terms of the two independent parameters, the black hole entropy is given by
\be
	 S_{\text{BH}} = \frac{\pi}{2 \Xi\? G_{\text{N}}^{4\rd}}\? \sqrt{\frac83\? b_0^3\? b_1 + 12 b_0^5\? b_1 + 12 b_0^4\? b_1^2 + b_0^6 (48 b_1^2-1) }\ .
\ee
Remarkably, we can rewrite the above formula in a more compact and suggestive form using the explicit relations between the conserved charges and the parameters $b_{0,1}$,
\be
 S_{\text{BH}} ( q_1 , q_2 , \cJ ) = \frac{\pi}{3 G_{\text{N}}^{4\rd}} \sqrt{\frac{108\? q_2^2 ( q_1 + q_2 ) + \cJ}{q_1 + 9 q_2}} \, .
\ee
Even if the latter formula seems more suggestive, we should stress that only two of the three charges are independent. In particular, one is not free to take the limit of vanishing $\cJ$ anymore and these black holes are always rotating, in contrast to the previously described twisted solutions.

\subsection{Universal Kerr-Newman-AdS black holes}
\label{Universal:KNAdS}

Apart from the solutions described above in the $T^3$ subtruncation, we can of course easily write down the analogous solutions in the truncation to minimal gauged supergravity.
Note that in terms of holographic meaning the latter solutions will \emph{not} be a subset of the former ones since the minimal truncation depends on the twist parameter $z$, \cf\,\eqref{V:minimal}.
This is in contrast with the static black holes above, which we could write down in full generality without specifying subtruncations, and, therefore automatically include the limit to the minimal theory.

We could in fact write down the most general thermal KN-AdS$_4$ black hole in minimal supergravity as written in \cite{Caldarelli:1998hg} but we directly focus on the supersymmetric limit in accordance with the preceding sections.
We give the conserved charges for completeness, leaving all details to the original reference.
The electric charge and angular momentum are related via
\be
	\cJ = - \frac{q}{2 g}\? \left( - 2 + \sqrt{4 + g^2 q^2} \right) ,
\ee
and the entropy reads
\be
 \label{SBH:KN:min}
 	S_{\text{BH}} ( q , \cJ ) = - \frac{2 \pi}{g G_{\text{N}}^{4\rd}} \? \frac{\cJ}{q} = \frac{\pi}{g^2 G^{4\rd}_{\text{N}}} \left( - 2 + \sqrt{ 4 + g^2 q^2} \right) ,
\ee
where $g = 2 L_{\text{AdS}_4}^{-1}$ with $L_{\text{AdS}_4}$ given in \eqref{V:minimal}.

In effect we already discussed the on-shell supergravity action of this solution in the minimal subtruncation in section \ref{sec:universal}, see also \cite{Bobev:2019zmz}.
One can recover it from the general form of \eqref{eq:NUTs} noting that the black holes exhibit two fixed points of the same orientation at the two poles of the sphere in the near-horizon geometry, $\epsilon_\text{SP} = \epsilon_\text{NP} = \epsilon$.
Upon an additional Legendre transform with respect to the electric charge $q$ and the angular momentum $\cJ$ one can obtain the Bekenstein-Hawking entropy \eqref{SBH:KN:min}.
We will capture this behavior in the next section where we discuss more generally the entropy function of all the black hole solutions in the matter-coupled theory using the gravitational blocks. 

\section{Gravitational blocks and black hole microstates}
\label{sec:gluing}

In this section we discuss the attractor mechanism for the class $\cF$ black holes and comment about the field theory interpretation of this result.
The attractor mechanism works as follows.
The Bekenstein-Hawking entropy $S_{\text{BH}} (p^i, Q_i , \cJ)$ of AdS black holes with arbitrary rotation and generic electric and magnetic charges is obtained by extremizing the functional \cite{Hosseini:2019iad}
\be
 \label{entropy:function}
 \cI (p^i , \cX^i , \omega) = \frac{\pi}{4 G_{\text{N}}^{4\rd}} \bigg( \cE (p^i , \cX^i , \omega) - 2 \ii \sum_{i = 1}^2 Q_i \cX^i - 2 \cJ \omega \bigg) + \mu \Big( 3 m \sum_{i = 1}^2 \cX^i - \ii \nu\? \omega - 2 \Big) \, ,
\ee
with respect to the chemical potentials $(\cX^i , \omega)$ conjugated to the conserved charges $(Q_i , \cJ)$, respectively, and the Lagrange multiplier $\mu$.
Extremizing \eqref{entropy:function} with respect to $\mu$ imposes a constraint among the chemical potentials that depends on the details of the model, $\nu = 0$ for the twisted black holes and $\nu = 1$ for the KN-AdS black holes.
Here, the functional
\be
 \cE (p^i , \cX^i , \omega) \equiv \sum_{r = 1}^2 \cB \big( X^1_{(r)}, X^2_{(r)}, \omega_{(r)} \big) \, ,
\ee
is obtained by gluing \emph{gravitational blocks}
\be
 \cB (X^i, \omega ) \equiv - \frac{F(X^i)}{\omega} \, ,
\ee
where $F(X^i) $ is the prepotential of 4d $\cN = 2$ gauged supergravity.
For \emph{twisted} AdS$_4$ black holes we use the $A$-gluing
\bea
 \label{A:gluing}
 X^i_{(1)} & = \cX^i - \ii \omega p^i \, , \qquad \omega_{(1)} = \omega \, , \\
 X^i_{(2)} & = \cX^i + \ii \omega p^i \, , \qquad \omega_{(1)} = - \omega \, ,
\eea
while for the KN-AdS black holes we use the identity gluing
\bea
 \label{id:gluing}
 X^i_{(1)} & = \cX^i - \ii \omega p^i \, , \qquad \omega_{(1)} = \omega \, , \\
 X^i_{(2)} & = \cX^i + \ii \omega p^i \, , \qquad \omega_{(1)} = \omega \,
.
\eea
Finally, the fixed attractor points $\mathring X^i_{(1)}$ and $\mathring X^i_{(2)}$ (critical point of \eqref{entropy:function})
are identified with the values of the supergravity sections $X^i$ at the South pole (SP) and the North pole (NP) of the sphere in the near horizon region.

For black holes of class $\cF$ the gravitational blocks are given by
\be
 \cB (X^1, X^2, \omega ) = \frac{\ii}{4 m\? \omega} \left( s^2 X^1 + s^1 X^2 \right) \sqrt{X^1 X^2} \, .
\ee
Notice that here we already used the effective form of the prepotential \eqref{eq:effectiveF} describing only the massless modes at the near horizon of our black holes.
We have therefore already imposed the vanishing of $k^\mathbb{R}_\Lambda X^\Lambda$, and thus eliminated $X^0$ (and $\cX^0$), in full analogy to the case discussed in \cite[Appendix A]{Hosseini:2020vgl}. We have also eliminated $A^0$ such that
\be
	P^\mathbb{R}_\Lambda \cX^\Lambda = 0\, , \qquad P^\mathbb{R}_\Lambda p^\Lambda = 0 \, ,
\ee
where we have suppressed the index $3$ for the moment maps, which is the only nonvanishing one.  This also leads to a redefinition of the electric charges associated with the two massless gauge fields $A^{1}$ and $A^2$ of the effective theory,
\be
 Q_1 \equiv q_1 + \frac{s^2}{4 m} \? q_0 \, , \qquad Q_2 \equiv q_2 + \frac{s^1}{4 m} \? q_0 \, .
\ee

\subsection{Twisted black holes}

For this class of black holes we use the $A$-gluing \eqref{A:gluing}.
We have checked that the entropy of static dyonic back holes with generic fluxes $(s^1, s^2)$ and rotating black holes of the $T^3$ model with $(s^1 = \frac1{3m}, s^2 = 0)$,
can be obtained by extremizing \eqref{entropy:function} with $\nu = 0$.
We have then verified that the values of the sections at the SP and the NP of the sphere in the near horizon region are given by
\bea
 X^i_{\text{SP, NP}} & = \frac{\ii}{2} \left( \mathring \cX^i \mp \ii \? \mathring \omega p^i \right) \, , \qquad i = 1 , 2
 \, .
\eea
The functional \eqref{entropy:function} has a natural field theory interpretation.
The five-dimensional $\cN=1$ Seiberg $\USp(2N)$ gauge theories wrapped on $\Sigma_\fg \times S^2$ are holographically dual to our black holes of class $\cF$ once they are uplifted to massive type IIA theory on $\Sigma_\fg \times_w S^4$.
Under some assumptions, at large $N$, their topologically twisted index on $\Sigma_{\fg} \times S^2 \times S^1$ was computed in \cite{Hosseini:2018uzp} and the final result can be written as
\be
 \label{twisted:index:5d}
 \log Z_{\text{twisted}} (\fs^i , \ft^i , \Delta_i) = - \frac{4}{27} \?F^{\text{Seiberg}}_{S^5} \sum_{i , j = 1}^2 \fs^i \ft^j \? \frac{\partial^2 ( \Delta_1 \Delta_2 )^{3/2}}{\partial \Delta_i \? \partial \Delta_j} \, ,
\ee
with $S^5$ free energy given in \eqref{F:S5} and
\be
 \label{CFT:constraint}
 \fs^1 + \fs^2 = 2 ( 1 - \fg ) \, , \qquad \ft^1 + \ft^2 = 2 \, , \qquad \Delta_1 + \Delta_2 = 2 \, .
\ee
Here, $(\fs^i , \ft^i)$ denote the magnetic fluxes through $(\Sigma_\fg, S^2)$, respectively, and $\Delta_{i}$, $i = 1,2$ are the chemical potentials that parameterize the Cartan of the $\SU(2)$ R-symmetry and the $\SU(2)$ flavor symmetry.
Define
\be
 \label{I:CFT:twisted}
 \cI_{\text{twisted}} (\Delta_i) \equiv \log Z_{\text{twisted}} (\fs^i , \ft^i , \Delta_i) - \ii \pi \sum_{i=1}^2 \wt Q_i \Delta_i \, .
\ee
The entropy of BPS static dyonic black holes can then be obtained as
\be
  S_{\text{BH}} (\fs^i , \ft^i, Q_i) = \cI_{\text{twisted}} (\Delta_i) \Big|_{\mathring \Delta_i} \, ,
\ee
where $\mathring \Delta_i$ is the extremum of $\cI_{(S^1 \times S^2 ) \times \Sigma_\fg} (\Delta_i)$.

Let us focus on static dyonic black holes where we know the large $N$ twisted index of the holographic dual field theory.
In the static limit $(\omega \to 0)$, the entropy function \eqref{entropy:function} is explicitly given by
\be
 \label{static:AdS:ent:func}
 \cI_{m\text{AdS}_4} ( p^i , \cX^i ) = \frac{\pi}{4 G_{\text{N}}^{4\rd}} \bigg( \frac{p^1 \cX^2 (3 s^2 \cX^1+ s^1 \cX^2) + p^2 \cX^1 ( s^2 \cX^1 + 3 s^1 \cX^2)}{4 m \sqrt{\cX^1 \cX^2}} - 2 \ii \sum_{i = 1}^2 Q_i \cX^i \bigg) \, .
\ee
The functional \eqref{static:AdS:ent:func} can be easily mapped to the field theory index \eqref{I:CFT:twisted} via \eqref{holo:dict}, \eqref{F:S5}, and 
\bea
 s^i  & = - \frac{\kappa}{6 m (1 - \fg)} \fs^i \, , \qquad && p^i  = - \frac{1}{6 m} \ft^i \, , \\
 Q_i & = 6 m G_{\text{N}}^{4\rd} \? \wt Q_i \, , && \cX^i = \frac{1}{3 m} \Delta_i \, , \qquad i = 1,2 \, .
\eea
This provides a microscopic derivation of the entropy of static \emph{dyonic} black holes in AdS$_4 \times \Sigma_\fg \times_w S^4$.

One can refine the index on $\Sigma_\fg \times S^2 \times S^1$ by adding a chemical potential $\omega$ for the angular momentum on $S^2$ \cite{Hosseini:2018uzp}.
Solving the refined index at large $N$ is a non-trivial problem. However, the gravitational picture suggests that it can be factorized into simpler building blocks as
\be
 \label{refined:conj}
 \log Z_{\text{twisted}} (\fs^i , \ft^i , \Delta_i | \omega) =
 - \frac{1}{2 \ii \omega } \left[ F^{\text{Seiberg}}_{S^3 \times \Sigma_\fg} \left( \fs^i , \Delta_i + \ii \frac{\omega}{2} \ft^i \right) - F^{\text{Seiberg}}_{S^3 \times \Sigma_\fg} \left( \fs^i , \Delta_i - \ii \frac{\omega}{2} \ft^i \right) \right] ,
\ee
where the $S^3 \times \Sigma_\fg$ free energy is given by \cite{Crichigno:2018adf}%
\footnote{One can swap between the conventions in \cite{Crichigno:2018adf} and here by setting $\fs_{1,2} = (1 - \fg) (1 \pm \hat \fn_M)$ and $\Delta_{1,2} = 1 \pm \wt\nu_{\text{AS}}$.}
\be
 \label{FS3:Sigmag}
 F^{\text{Seiberg}}_{S^3 \times \Sigma_\fg} ( \fs^i , \Delta_i ) = - \frac{8 \sqrt{2} \? \pi}{15} \frac{N^{5/2}}{\sqrt{8 - N_f}} \sum_{i = 1}^2 \fs^i \frac{\pd ( \Delta_1 \Delta_2 )^{3/2}}{\pd \Delta_i} \, .
\ee
It would be interesting to generalize the computations of \cite{Hosseini:2018uzp} to the refined case and prove the above conjecture. We naturally expect \eqref{refined:conj} to hold for other classes of 5d SCFTs with a holographic dual.

\subsection{Kerr-Newman-AdS black holes}

For this class of black holes we use the identity gluing \eqref{id:gluing}. We have checked that the entropy of the KN-AdS black holes of the $T^3$ model can be obtained by extremizing \eqref{entropy:function} with $\nu = 1$.%
\footnote{The entropy function for the universal KN-AdS$_4$ black holes discussed in section \ref{Universal:KNAdS} (sticking to the normalization of $F_{\text{min}}$ in \eqref{prepotential:minimal}) is given by $\displaystyle \cI_{\text{min}} ( \wt \cX , \omega) = \frac{\pi}{4 G_{\text{N}}^{4\rd}} \left( \cE - 2 \ii \wt \cX q - 2  \omega \cJ \right) + \mu ( g \wt \cX - \ii \omega - 2 ) .$}
Moreover, we have verified that the solutions to the saddle point equations, denoted by $(\mathring \cX^i , \mathring \omega)$, are related to the values of the sections at the SP and the NP of the sphere by
\bea
 \bar X^i_{\text{SP}} & = - \frac{\ii}{2} \? \mathring X^{i}_{(1)}  = - \frac{\ii}{2} \Big( \mathring\cX^i - \ii \mathring \omega p^i \Big) \, , \\
 X^i_{\text{NP}} & = -\frac{\ii}{2}  \mathring X^{i}_{(2)} = - \frac{\ii}{2} \Big( \mathring\cX^i + \ii \mathring \omega p^i \Big) \, , \qquad i = 1 , 2 \, ,
\eea
for $i = 1, 2$. Notice that here the values of the sections can be identified with the critical values of the gluing quantities $X^{i}_{(r)}$ up to a complex conjugate. Even though the general gluing rules allow for magnetic charges, the explicit solutions we could write down above in the $T^3$ and minimal subtruncations have vanishing magnetic charges due to additional regularity constraints. In these cases the entropy function simplifies and agrees with the one proposed in \cite{Choi:2018fdc} and rederived in \cite{Cassani:2019mms} for Kerr-Newman black holes with AdS$_4 \times S^7$ asymptotics.

The functional \eqref{entropy:function} for dyonic KN-AdS black holes with generic fluxes $(s^1 , s^2)$ can be interpreted as the Legendre transform of the large $N$ partition function of the holographic dual field theory on $(S^1 \times S^2) \times \Sigma_\fg$. Here, the $S^2$ is \emph{not} twisted while there still exists a partial topological twist on $\Sigma_\fg$.
This is an intriguing example of a five-dimensional partition function that from one angle can be thought of as a 3d topologically twisted index on $\Sigma_\fg \times S^1$ (with a KK tower of modes on $S^2$)
and from another angle as a 3d superconformal index (SCI) on $S^2 \times S^1$ (with a KK tower of modes on $\Sigma_\fg$).
It would certainly be interesting to derive this partition function using localization and study its large $N$ behavior.
The gravitational viewpoint suggests the factorization of this partition function, at least in the large $N$ limit, as
\be
\label{eq:sci:conj}
 \log Z_{\text{SCI}} (\fs^i , \ft^i , \Delta_i | \omega) =
 - \frac{1}{2 \ii \omega } \left[ F^{\text{Seiberg}}_{S^3 \times \Sigma_\fg} \left( \fs^i , \Delta_i + \ii \frac{\omega}{2} \ft^i \right) + F^{\text{Seiberg}}_{S^3 \times \Sigma_\fg} \left( \fs^i ,  \Delta_i - \ii \frac{\omega}{2} \ft^i \right) \right] ,
\ee
where $F^{\text{Seiberg}}_{S^3 \times \Sigma_\fg} (\fs^i , \Delta_i)$ is given in \eqref{FS3:Sigmag}. We again expect \eqref{eq:sci:conj} to hold for other classes of 5d SCFTs with a holographic dual.

\section{Discussion and outlook}
\label{sec:conclusions}

A major motivation for this work stems from the rather enigmatic nature of the class $\cF$ theories, obtained by twisted compactification of 5d $\cN = 1$ CFTs on a generic Riemann surface.
The bulk duals of such SCFTs interpolate between AdS$_6$ and AdS$_4 \times \Sigma_\fg$ vacua, leading to a natural holographic interpretation as RG flows across dimensions.
The AdS$_4$ region was explicitly constructed in massive type IIA in \cite{Bah:2018lyv} based on the general classification of \cite{Passias:2018zlm}.
Due to the construction here we also expect that the type IIB uplift via \cite{Malek:2019ucd} will fit in the general classification of \cite{Passias:2017yke}. Such bulk constructions  indicate the existence of an interacting conformal phase in the IR limit of the \emph{parent} 5d SCFTs on $\Sigma_\fg$.

The parent 5d SCFTs are strongly coupled microscopic theories and
can be realized at the intersection of $N$ D4-branes, $N_f$ D8-branes and orientifold planes in the massive IIA case,
or alternatively by utilizing $(p,q)$ 5-branes in type IIB.
Various supersymmetric observables for such CFTs have been computed at large $N$, see for example \cite{Jafferis:2012iv,Chang:2017mxc,Fluder:2018chf,Hosseini:2018uzp,Crichigno:2018adf,Fluder:2019szh,Uhlemann:2019ypp},
and they exhibit the $N^{5/2}$ and $N^4$ scaling of the number of degrees of freedom of the D4-D8-O8 system and the $(p,q)$ 5-branes, respectively.
It would be desirable to have a three-dimensional description of this class of SCFTs, similar to the class $\cS$ program where Gaiotto \cite{Gaiotto:2009we} identified a family of isolated $\cN = 2$ SCFTs,
describing the theory of $N$ coincident M5-branes wrapped on a Riemann surface.

Another interesting research direction concerns the correspondences \ala\,AGT \cite{Alday:2009aq}.
The basic premise is that the compactification of 6d $\cN = (2,0)$ theory on $S^4$ leads to a two-dimensional Toda CFT on a Riemann surface, dual to the 4d supersymmetric gauge theory obtained by reduction on the surface.
It would be intriguing to derive the analogous 3d-2d correspondence for the class $\cF$ theories and exploit it to understand the microscopic origin of the entropy of the black holes presented in this paper.
This line of thought has proven to be useful, see \eg\,\cite{Gang:2014qla} and references thereto, in studying the microscopic free energy of various bulk solutions arising from M5-branes wrapped on hyperbolic 3-manifolds \cite{Donos:2010ax}.

There are also plenty of open problems in supergravity, which seem more immediately reachable.
Let us mention some of them.
\begin{itemize}
	\item[--] The proof that our starting point, the 6d F(4) supergravity coupled to an abelian vector multiplet, comes from a consistent truncation of massive IIA on $S^4$ is strictly speaking still lacking.
	Such a reduction was performed on the abelian T-dual solution in \cite{Malek:2019ucd} and we presented additional arguments why it must be true.
	It seems very likely that this problem can be completely settled with the exceptional field theory (EFT) techniques \cite{Hohm:2013pua,Malek:2019ucd}.
	It would be further very interesting if EFT techniques can be used to determine the full KK spectra for the holographic backgrounds discussed here, in analogy to \cite{Malek:2019eaz,Malek:2020yue,Varela:2020wty}.

	\item[--] An interesting generalization is to enlarge the reduction ansatz to include the symmetry group of the Riemann surface.
	In particular, we need to include extra gauge fields arising from the six-dimensional metric, which would give rise to additional four-dimensional vector multiplets. Such a bulk construction would be the analogue of equivariant integration, see \eg\,\cite{Hosseini:2020vgl}, in the holographic dual field theory.
	A related observation is that one can also add punctures to the Riemann surface for the class $\cS$ theories \cite{Gaiotto:2009we,Gaiotto:2009hg}
	and their bulk duals \cite{Gaiotto:2009gz,Bah:2017wxp,Bobev:2019ore}, while preserving some amount of supersymmetry.
	It would be interesting to understand better the analogous story for the class $\cF$ theories.

	\item[--] We already discussed more extensively the squashed sphere and black hole solutions in AdS$_4$ arising from the reduced model.
	We should note the existence of other purely Euclidean solutions such as \cite{Freedman:2013oja} and more recently \cite{Bobev:2020pjk}.
	Both types of solutions can be directly embedded in the $T^3$ subtruncation of our model and have an intriguing holographic interpretation.
	It would be also interesting to generalize these solutions to the full model with hypermultiplet gauging.

	\item[--] An interesting further extension is to start from a bigger 6d theory with even more matter multiplets.
	Some very special examples can be found as arising from type IIB compactifications \cite{Malek:2019ucd}.
	The reduction of the 6d F(4) supergravity coupled to \emph{two} vector multiplets on a Riemann surface gives rise to an extra vector multiplet in the 4d F(4) supergravity constructed here.
	It is tempting to speculate that this completes the square root prepotential \eqref{F:incomplete} to the $STU$ model,
	\be
	 F = - \ii \? \sqrt{X^0 X^1 X^2 X^3} \, .
	\ee
	A further massless subtruncation might then lead to the $S T^2$ model generalizing the $T^3$ truncation here.

	\item[--] An open avenue for exploration is the possibility of adding higher-derivative corrections to the effective supergravity model.
	It is worth noting that the structure of the higher-derivative theory is much better studied in four dimensions \cite{Bergshoeff:1980is,deWit:2010za,Butter:2013lta}
	and via the uplift on $\Sigma_\fg$ we can hope to understand better such corrections also in six dimensions.
	Alternatively, one can also adopt a more practical approach and try to directly fix the 		higher-derivative corrections in 4d via holography, see \cite{Bobev:2020egg}.

\end{itemize}

\section*{Acknowledgements}

The authors would like to thank Achilleas Passias for fruitful discussions and collaboration in the early stages of this work, and Alberto Zaffaroni for useful discussions and comments.
SMH is supported in part by WPI Initiative, MEXT, Japan at IPMU, the University of Tokyo, JSPS KAKENHI Grant-in-Aid (Wakate-A), No.17H04837 and JSPS KAKENHI Grant-in-Aid (Early-Career Scientists), No.20K14462.
KH is supported in part by the Bulgarian NSF grants DN08/3, N28/5, and KP-06-N 38/11.

\bibliographystyle{ytphys}

\bibliography{classF}

\end{document}